**Signatures of unconventional superconductivity near reentrant and fractional quantum anomalous Hall insulators**

Fan Xu[1,2†], Zheng Sun[1†], Jiayi Li[1†], Ce Zheng[3†], Cheng Xu[4], Jingjing Gao[5], Tongtong Jia[1,2], Yanfei Su[6], Kenji Watanabe[7], Takashi Taniguchi[8], Bingbing Tong[9], Li Lu[9,10], Jinfeng Jia[1,2,6,10,11], Zhiwen Shi[1], Shengwei Jiang[1], Junhao Lin[6,11], Yuanbo Zhang[5,10], Yang Zhang[4,12], Shiming Lei[3*], Xiaoxue Liu[1,2,10*], and Tingxin Li[1,2,10*]

[1]State Key Laboratory of Micro-nano Engineering Science, Tsung-Dao Lee Institute, Shanghai Jiao Tong University, Shanghai, China

[2]Key Laboratory of Artificial Structures and Quantum Control (Ministry of Education), School of Physics and Astronomy, Shanghai Jiao Tong University, Shanghai, China

[3]Department of Physics, Hong Kong University of Science and Technology, Clear Water Bay, Hong Kong SAR, China

[4]Department of Physics and Astronomy, University of Tennessee, Knoxville, TN, USA

[5]State Key Laboratory of Surface Physics and Department of Physics, Fudan University, Shanghai, China

[6]State Key Laboratory of Quantum Functional Materials, Department of Physics and Guangdong Basic Research Center of Excellence for Quantum Science, Southern University of Science and Technology (SUSTech), Shenzhen 518055, China

[7]Research Center for Electronic and Optical Materials, National Institute for Materials Science, 1-1 Namiki, Tsukuba, Japan

[8]Research Center for Materials Nanoarchitectonics, National Institute for Materials Science, 1-1 Namiki, Tsukuba, Japan

[9]Beijing National Laboratory for Condensed Matter Physics and Institute of Physics, Chinese Academy of Sciences, Beijing, China

[10]Hefei National Laboratory, Hefei, China

[11]Quantum Science Center of Guangdong-Hong Kong-Macao Greater Bay Area (Guangdong), Shenzhen 518045, China

[12]Min H. Kao Department of Electrical Engineering and Computer Science, University of Tennessee, Knoxville, Tennessee, USA

[†]These authors contributed equally to this work.

[*]Emails: phslei@ust.hk, xxliu90@sjtu.edu.cn, txli89@sjtu.edu.cn

**Abstract:**

Two-dimensional moiré Chern bands provide an exceptional platform for exploring a variety of many-body quantum phases at zero magnetic field within a lattice system. One particular intriguing possibility is that flat Chern bands can, in principle, support exotic superconducting phases together with fractional topological phases. Here, we report the observation of integer and fractional quantum anomalous Hall effects, the reentrant quantum anomalous Hall effect, and superconductivity within the first moiré Chern band of twisted bilayer $MoTe_2$. The superconducting phase emerges from a normal state exhibiting anomalous Hall effects. Our results present the first example of superconductivity emerging within a flat Chern band that simultaneously hosts fractional quantum anomalous effects, a phenomenon never observed in any other systems. Our work expands the understanding of emergent quantum phenomena in moiré Chern bands, and offers a nearly ideal platform for engineering Majorana and parafermion zero modes in gate-controlled hybrid devices.

**Main:**

The quantum geometry of electron wavefunctions, encompassing Berry curvature and quantum metric, profoundly influences the emergence and properties of various quantum states. For example, the recently observed fractional quantum anomalous Hall (FQAH) effect (*1-13*) (i.e. fractional Chern insulators (*14-18*) at zero magnetic field) occurs when the flat Chern band is close to the ideal condition for quantum geometry. Although FQAH effects have been experimentally realized in twisted bilayer $MoTe_2$ ($tMoTe_2$) (*1-4, 6-10*), the competition between FQAH effects and other interaction-driven quantum states within the moiré Chern band, is still largely unexplored. On the other hand, the quantum metric is believed to play an important role in facilitating superconductivity within a topological flat band (*19-23*). Two-dimensional (2D) flat band superconductors have attracted great attention since the discovery of superconductivity in twisted bilayer graphene (*24-27*). Since then, superconductivities have been observed in twisted multilayer graphene (*28-31*), rhombohedral graphene moiré systems (*13*, *32*), moiré-less crystalline graphene (*33-39*), and most recently, in twisted bilayer $WSe_2$ ($tWSe_2$) (*40*, *41*). However, the coexistence of FQAH effects and superconductivity within a single topological flat band remains elusive in experiments.

In this work, we report transport studies of the first moiré Chern band of ~3.8°-4° $tMoTe_2$, with significantly improved device quality compared to previous studies (*3*,*4*), enabling the observation of intriguing new quantum states. Specifically, we observe integer quantum anomalous Hall (IQAH) effect at moiré hole filling factor of $\nu_h = 1$, fractional Chern insulators at $\nu_h$ = 2/3, 3/5, 4/7, 5/9, and eight topologically trivial correlated insulators. In between of $\nu_h$ = 2/3 and 3/5 FQAH states, a reentrant integer quantum anomalous Hall (RIQAH) state – the zero magnetic field analogue of reentrant quantum Hall states (*42-47*) – emerges at temperature below 300 mK. Remarkably, signatures of unconventional superconductivity have been observed, neighboring with the FQAH states and RIQAH states. The maximum onset superconducting transition temperature is approximately 1.2 K, with the zero-resistance state achieved around 300 mK, and a perpendicular critical magnetic field as high as approximately 0.8 T. Notably, the normal state of the observed superconductivity exhibits anomalous Hall (AH) effect and magnetic hysteresis. These distinctive behaviors differentiate it from prior moiré superconductors and underscore its unconventional origin.

**Transport phase diagram of high quality $tMoTe_2$ devices**

To improve device quality, we implemented two key improvements. First, high-quality $2H\text{-}MoTe_2$ single crystals were grown using the flux method (*48*), resulting in reduced defect densities and improved carrier mobility (fig. S1 in *48*). Second, although the twisted $MoTe_2$ devices were fabricated largely following the procedures described in our previous studies (*4*, *9*), an additional annealing step was introduced after stack assembly to further improve the moiré uniformity and contact quality (*48*). Figure 1A shows the schematic of the device structure. The top and bottom graphite gates are used to independently control the $\nu_h$ and the vertical electric displacement field $D$. The twist angle $\theta$ of the measured devices are calibrated through quantum oscillations observed under perpendicular magnetic fields (fig. S2). We measured six devices in dilution refrigerators with $\theta \approx 3.8°\text{-}4.0°$ (see Table 1 in *48* for device summary). Below, we first focus on Device 1, which exhibits the highest device quality among them. The twist angle of Device 1 is calibrated as 3.83°±0.02°, corresponding to a moiré density $n_M \approx (4.16\pm0.05)\times10^{12}$ cm$^{-2}$ and a moiré lattice constant $a_M \approx 5.27\pm0.03$ nm.

Figure 1C shows the longitudinal resistivity $\rho_{xx}$ of Device 1 as a function of $D$ and $\nu_h$ at temperature $T$ = 2 K and zero magnetic field. At $\nu_h$ = 1, 2/3, and 3/5, local dips in $\rho_{xx}$ can be identified around $D$ = 0, corresponding to the IQAH state at $\nu_h$ = 1 and FQAH states at $\nu_h$ = 2/3, 3/5, consistent with previous studies (*3*, *4*). The applied $D$-field can drive the quantum anomalous Hall states into topologically trivial, correlated insulating states at $\nu_h$ = 1 and 2/3. Additionally, correlated insulators are also observed at $\nu_h$ = 1/2, 1/3, 1/4, and 1/8, under finite $D$-fields. Figure 1D shows a similar $\rho_{xx}$-$\nu_h$-$D$ map measured at a lower temperature, $T$ = 900 mK. The main features remain similar to those observed at $T$ = 2 K, but become more pronounced. Additionally, around $D$ = 0, $\rho_{xx}$ dips associated with the $\nu_h$ = 4/7 FQAH state and $\rho_{xx}$ peaks of correlated insulators at $\nu_h$ = 3/2 and 4/3 begin to develop at $T$ = 900 mK, becoming more prominent at lower temperatures (Fig. 1E). These states were not observed in previous transport studies (*3*, *4*, *9*, *10*).

Figure 1, E and F, further show $\rho_{xx}$ and Hall resistivity $\rho_{xy}$, respectively, as a function of $\nu_h$ and $D$ at $T = 100$ mK. $\rho_{xx}$ and $\rho_{xy}$ have been symmetrized and anti-symmetrized using data measured under perpendicular magnetic field $B = \pm 0.1$ T. In comparison to the 900 mK $\rho_{xx}$-$\nu_h$-$D$ map, several intriguing new features emerge at ultra-low temperatures: 1) Between the $\nu_h = 2/3$ and 3/5 FQAH states, another state emerges, characterized by vanishing $\rho_{xx}$ and large $\rho_{xy}$. As we elaborate in the following, this state is a RIQAH state. 2) Another RIQAH state occurs at $\nu_h$ slightly higher than 2/3 under a finite $D$-field. 3) At $\nu_h$ from approximately 0.71 to 0.76 around $D = 0$, both $\rho_{xx}$ and $\rho_{xy}$ vanish, suggesting the emergence of a superconducting state. These features will be the main focus of the following sections.

**IQAH, FQAH and RIQAH effects**

Figure 2, C through H, shows $\rho_{xx}$ and $\rho_{xy}$, respectively, at different $\nu_h$, $D$ points with local $\rho_{xx}$ dips and nearly quantized $\rho_{xy}$ (marked in Fig. 2, A and B), as a function of $B$ at a nominal mixing chamber temperature $T_{MC}$ of 15 mK (*48*). Temperature dependence data at these $\nu_h$, $D$ points are shown in fig. S3. Clear magnetic hysteresis loops are observed at all selected $\nu_h$, $D$ points. The coercive field is largest at $\nu_h = 1$ and $D = 0$, around 100 mT, and approximately 15-30 mT for other $\nu_h$, $D$ points at $T_{MC} = 15$ mK. As shown in Fig. 2C, at $\nu_h = 1$ and $D = 0$, $\rho_{xy}$ is quantized at $h/e^2$ at zero magnetic field, consistent with previous studies (*3*, *4*). Notably, at $\nu_h = 2/3$ and $D = 0$ (Fig. 2F), $\rho_{xy}$ is quantized at $3h/2e^2$ with a residual $\rho_{xx}$ below 300 Ω at zero magnetic field. At $\nu_h = 3/5$ (Fig. 2G) and 4/7 (Fig. 2H) at $D = 0$, the zero-field AH signals also approach to the expected quantized value, but with less accuracy and relatively large residual $\rho_{xx}$ (~ 5 kΩ), as compared to $\nu_h = 2/3$. This might be attributed to less good contacts at low filling factors and smaller FQAH gaps at $\nu_h = 3/5$ and 4/7.

Figure 2D illustrates similar $B$-scan data of $\rho_{xx}$ and $\rho_{xy}$ at $\nu_h = 0.63$ and $D = 0$. Intriguingly, at this incommensurate moiré filling, the measured $\rho_{xy}$ is also quantized at $h/e^2$ at $B = 0$, with residual $\rho_{xx} < 1$ kΩ. We refer to this state as the RIQAH effect, the lattice analogue of the reentrant integer quantum Hall (RIQH) effect observed previously in Landau levels (*43-47*). Similarly, at $\nu_h = 0.70$ and $D = \pm 27$ mV/nm, another RIQAH appears (Fig. 2E). Although it emerges at a finite $D$ field at $B = 0$, it extends across a broader $D$ field range under higher magnetic fields (fig. S4). Similar RIQAH states are also observed in Device 2-6, with $\theta$ ranging from ~ 3.8° to 4°.

The nature of the IQAH, FQAH, and RIQAH states can be further examined by measuring $\rho_{xx}$ as a function of $\nu_h$ and $B$. As shown in Fig. 2, I and J (also in figs. S5, S7, S10, S13, S14), all the states mentioned above featured by local $\rho_{xx}$ dips which show linear shift in $\nu_h$ with increasing $B$. The dispersion of the FQAH states agrees well with expected Chern number $C$ according to the Streda formula $n_M \frac{d\nu_h}{dB} = C\frac{e}{h}$, as illustrated by the dashed lines. In particular, the $\nu_h$ = 2/3, 3/5, and 4/7 FQAH states persist to $B$ = 0 T, while the $\nu_h$ = 5/9 state persists to $B \approx 0.5$ T. At $T_{MC}$ =15 mK, the plateau of $\nu_h$ = 1 IQAH state spans over a large filling range ($\nu_h \sim 0.9$ to 1.25), making it difficult to determine $C$ based on the Streda formula. Instead, we fit $C$ of the $\nu_h$ = 1 IQAH state at higher temperature, which is in good agreement with the expected value of $C$ = 1 (fig. S5). However, for the RIQAH states, although $\rho_{xy}$ shows a clear $h/e^2$ plateau, the measured dispersion does not correspond to $C$ = 1; instead, it is closer to its filling factors (figs. S5, S10, S13, S14). The underlying mechanism requires further theoretical and experimental investigations.

Figure 2K demonstrates the symmetrized $\rho_{xx}$ and anti-symmetrized $\rho_{xy}$ measured under $B = \pm 0.1$ T as a function of $\nu_h$, at $T_{MC}$ = 15 mK and $D$ = 0. The quantized Hall plateaus and corresponding local dips in $\rho_{xx}$ for the $\nu_h$ = 1 IQAH state, $\nu_h$ = 2/3, 3/5, 4/7 FQAH states, and the RIQAH states are clearly observed. Similar features can also be identified in $\sigma_{xx}$ and $\sigma_{xy}$ plots, as shown in fig. S6 (*48*). At $T_{MC}$ = 15 mK, the quantization of the $\nu_h$ = 2/3 FQAH state is within 0.5% accuracy of $2e^2/3h$ over a narrow range of $\nu_h$ = 2/3 ± 0.005, and the quantization of the RIQAH state is within 0.5% accuracy of $e^2/h$ in the range of $\nu_h$ = 0.63 ± 0.007. The quantization is less accurate for the 3/5 and 4/7 FQAH states, with deviations of about 3% and 1.5%, respectively, from $3e^2/5h$ and $4e^2/7h$.

These data solve an important puzzle regarding the dip in $\rho_{xy}$ between $\nu_h$ = 2/3 and 3/5 FQAH states observed in previous studies of ~3.7°- 4° $tMoTe_2$ (*3*, *4*). As shown in the temperature dependence data of $\rho_{xx}$ and $\rho_{xy}$ versus $\nu_h$ (Fig. 3, A and B, fig. S11), the RIQAH state has a smaller energy scale than the $\nu_h$ = 2/3 FQAH state, therefore requires lower temperatures (< 300 mK) and reduced disorder broadening for reaching quantization. At higher temperatures, the measured $\rho_{xy}$ and $\rho_{xx}$ around $\nu_h$ = 2/3 are essentially the same as previous reported results. Note that in between of $\nu_h$ = 3/5 and 4/7, similar $\rho_{xy}$ dip has also been observed at low temperatures, which may indicate the presence of more RIQAH states with an even smaller energy scale.

In two-dimensional electron gas system under strong magnetic fields, the RIQH state is one type of quantum solid that competes with the fractional quantum Hall liquid (*42-47*). The RIQH effect occurs at fractional fillings of Landau levels, and can be understood as the formation of interaction-driven triangular Wigner crystal (WC) or electron bubble phases (where several electrons/holes become localized near each WC lattice site), superimposed on fully-filled Landau levels. In the lowest Landau level, previous studies of $Al_xGa_{1-x}As/Al_{0.32}Ga_{0.68}As$ heterostructures reported (*46*) a RIQH state between Landau level filling factor $\nu_{LL} = 2/3$ and 3/5, also centered at $\nu_{LL} \approx 0.63$. Our results represent the zero-field version of this effect for the first time.

It is worth noting that the observed RIQAH effects are distinct from topological charge density waves or generalized anomalous Hall crystals. These states have been observed in various moiré systems (*49-52*), also characterized by integer quantized AH signals at fractional filling factors. However, these states occur exclusively at commensurate moiré fillings, which can be understood in terms of interaction-induced enlarging of the moiré unit cell. In contrast, the RIQAH state does not necessarily occur at commensurate moiré fillings. The RIQAH state also differs from the recent observed extended quantum anomalous Hall (EQAH) effect in rhombohedral graphene moiré systems (*12*), where integer quantized AH conductance persists over a broad moiré filling factor range. Moreover, the EQAH state even suppresses some FQAH states at sufficiently low temperatures and bias voltages.

**Signatures of unconventional superconductivity**

Next, we focus on an oval-shaped region in the $\nu_h$ -$D$ map, as highlighted by the black dashed lines in Fig. 2, A and B. Remarkably, this region exhibits both vanishing $\rho_{xx}$ and vanishing $\rho_{xy}$, which is substantially different from the IQAH, FQAH and RIQAH states with vanishing $\rho_{xx}$ and quantized anomalous $\rho_{xy}$. As shown in Fig. 3, A-C, between the $\nu_h = 1$ IQAH state and the $\nu_h = 2/3$ FQAH state, most fillings at $D = 0$ exhibit non-quantized but finite AH signals, accompanied by weak temperature dependence in $\rho_{xx}$, corresponding to an AH metal phase as reported previously (*3*, *4*). Notably, $\rho_{xy}$ shows a sudden drop between $\nu_h$ around 0.71 to 0.76, reaching zero below approximately 300 mK (Fig. 3C). At higher temperatures, the AH signal gradually recovered in this region. Correspondingly, $\rho_{xx}$ in the same filling range also drops rapidly with decreasing temperature. Under $B = \pm 0.1$ T, $\rho_{xx}$ saturates at approximately 300 Ω at low temperatures (Fig. 3A), while at $B = 0$ T, it reaches zero within the measurement noise floor below approximately 300 mK (Fig. 3D). Figure 3E shows the map of $\rho_{xx}$ versus $T$ and $\nu_h$, revealing a superconducting dome-like feature. We find that the observed superconducting states are highly sensitive to both the twist angle and the device quality. Signatures of superconductivity were observed in three high-quality devices with $\theta \approx 3.83°$-$3.93°$ (figs. S10-S12).

The $\rho_{xx}$-$\nu_h$-$B$ maps shown in Fig. 2, I and J, provide further evidences for the existence of superconductivity in the moiré Chern band of $tMoTe_2$. Unlike the IQAH, FQAH and RIQAH states, which exhibits a dispersion with $B$ in $\nu_h$ as mentioned above, the state between $\nu_h$ around 0.71 to 0.76 does not disperse with $B$, and can be suppressed by a moderate magnetic field. At higher magnetic fields ($B > \sim 2$ T), another reentrant Chern state emerges at $\nu_h > 2/3$, separated from the superconducting state by a resistive state. All of these observations are reproducible in another pair of contacts (fig. S7) as well as in another device (fig. S10). As shown in the $B$-dependent and $T$-dependent $\nu_h$-$D$ maps of $\rho_{xx}$ and $\rho_{xy}$ (figs. S4, S9, S10, S11), superconductivity is the most pronounced at $D = 0$ but extends into finite $D$, surrounded by the AH metal and RIQAH phases.

As illustrated in Fig. 3F, the superconducting transition is relatively broad, with a maximum onset transition temperature $T_{onset}$ of approximately 1.2 K, eventually reaching the zero-resistance state at about 300 mK. The measured differential resistance d$V$/d$I$ as functions of dc current $I_{dc}$ and $B$ are shown in fig. S8 (*48*). Obvious non-linearity can be observed until the superconducting critical magnetic field $B_c \sim 0.6$ T-0.8 T is reached. The d$V$/d$I$-$I_{dc}$ does not exhibit sharp coherence peaks, consistent with the broad superconducting transition, which is likely limited by disorder effects. Figure S11 also shows the direct $I$-$V$ measurements of the superconducting states in device 2, where clear nonlinear superconducting $I$–$V$ characteristics are observed below ~ 1 K, with a critical current of approximately 5-10 nA.

Figure 4, A-G, shows the $B$-field scan data (within ±75 mT) of $\rho_{xx}$ and $\rho_{xy}$ in the superconducting region at $D = 0$ and $\nu_h = 0.74$ of Device 1. At low temperatures, the zero-resistance state persists within a small magnetic field range, approximately ± 6 mT at 75 mK. Beyond this range, $\rho_{xx}$ begins to increase slowly with increasing $B$, but it remains significantly lower than the normal-state resistivity within the scanned $B$-field range, presumably due to the unbinding vortices (*48*). Magnetic switching behavior is observed, featured by two sharp peaks. In the same $B$-field range, $\rho_{xy}$ stays at zero, except at the magnetic switch peaks (Fig. 4, B and C). At higher temperatures, both $\rho_{xx}$ and $\rho_{xy}$ deviate from zero resistance, and clear magnetic hysteresis loops, along with AH effects at $B = 0$, can be observed (Fig. 4, D-G). Figure 4I illustrates $\rho_{xx}$ as a function of $T$ under different $B$, showing that strong magneto-resistance occurs only below $T_{onset}$. As for $\rho_{xy}$, it exhibits strong $T$-dependence but weak $B$-dependence in the small magnetic field limit, while becomes nearly temperature-independent when $B > B_c$, as shown in Fig. 4J. The Ginzburg-Landau superconducting coherence length $\xi$ is estimated to be $\xi \approx$ 20-23 nm (*48*). The averaging distance between holes at $\nu_h = 0.73$ for 3.83° $tMoTe_2$ is $d_{hole} \approx 1/\sqrt{\nu_h n_M} \approx$ 6 nm, leading to $\frac{\xi}{d_{hole}} \approx$ 3.3-3.8, suggesting a strong coupling of Cooper pairing.

The observed magnetic switching behaviors and normal-state AH effects are unusual for a superconductor. We note these peculiar behaviors align with the recent observed signatures of chiral superconductivity in rhombohedral multilayer graphene (*39*). In that system, a superconducting state emerges within a spin- and valley-polarized quarter-metal phase, exhibiting similar magnetic switching behaviors and normal-state AH effects as we observed. In $tMoTe_2$, emergent ferromagnetism has been proposed to arise from interaction-induced valley polarization (*53-58*), and it is evident in optical measurements of 3.5° to 4° $tMoTe_2$, across a wide range of $\nu_h$ from approximately 0.4 to 1.2 (*1*, *59*). Consequently, the observed superconductivity in $tMoTe_2$ likely develops from a (partially) spin- and valley-polarized Fermi surface, which is consistent with the AH effects observed in the normal state of superconductivity.

In $tWSe_2$, superconductivity (*40*, *41*) was observed in devices with relatively large $\theta$, where ferromagnetism is absent due to the weakened correlation effects. The observed superconductivity in $tWSe_2$ emerges around $\nu_h = 1$ and is most prominent at finite electric fields, consistent with the feature of van Hove singularities (VHS). The relatively small $B_c$ (~0.1 T) in $tWSe_2$ further suggests that the pairing is likely of intervalley nature. In contrast, single-particle band calculations for ~3.9° $tMoTe_2$ reveal a VHS at $\nu_h \approx 0.8$ at $D = 0$, which shifts to larger $\nu_h$ with enhanced peaks at higher $D$ (fig. S16), resembling the behavior in $tWSe_2$. Experimentally, however, the superconducting states in ~3.83°-3.93° $tMoTe_2$ are centered around $D = 0$ and exhibit a relatively large $B_c$. Moreover, the observed VHS features deviate substantially from the single-particle prediction. In particular, two parabola-shaped VHS branches are resolved near $\nu_h \approx 0.7$ and $\nu_h \approx 1.3$ at $D = 0$ (fig. S16), whose envelope closely follows the boundary of the zero-field anomalous Hall (ferromagnetic) region. This splitting is naturally explained by interaction-induced valley polarization, and the coincidence of the two VHS branches with the ferromagnetic boundary suggests that the spontaneous valley polarization is closely connected to the diverging density of states associated with the VHS. These observations suggest that the interplay of interaction-renormalized VHS and ideal quantum geometry may play a significant role in facilitating superconductivity in 3.83°-3.93° $tMoTe_2$, potentially paving the way for intravalley pairing.

**Discussions and conclusions**

The RIQAH states, FQAH states and superconductivity exhibit subtle competitions within the first moiré Chern band in $tMoTe_2$. For $\nu_h < 2/3$, the RIQAH solid at $\nu_h \approx 0.63$ competes with the $\nu_h = 2/3$ FQAH liquid, resembling the physics of the lowest Landau level in two-dimensional electron gas with short-range disorders (*46*). Additionally, in the lowest Landau level, pinned WC dominate over the fractional quantum Hall states for $\nu_{LL} < 1/3$ or $\nu_{LL} < 1/5$, depending on the details of disorders, effective mass, and other sample parameters (*60-62*). The continuously insulating region from $\nu_h = 0$ to ~0.4 observed in $tMoTe_2$ (Fig. 1C) may suggest the formation of WC at zero magnetic field, which requires further theoretical and experimental investigations. On the other hand, for $2/3 < \nu_h < 1$, the quantum states formed in the first moiré Chern band of ~3.8°-4° $tMoTe_2$ fundamentally differ from those in the lowest Landau level at $2/3 < \nu_{LL} < 1$. Both the RIQAH states at $\nu_h \approx 0.7$ and the superconducting states are absent in the lowest Landau level. The RIQAH state also seems to compete with the superconducting state. In Device 1-3, the $\nu_h \approx 0.7$ RIQAH states take over the superconducting region under high magnetic fields (figs. S4, S12). Moreover, in devices with slightly different $\theta$ or stronger disorders (Device 4-6), the RIQAH state also tends to replace the superconducting states (figs. S13-S15).

Theoretically, anyon superconductivity was proposed in the fractional quantum Hall regime (*63-67*). In a flat Chern band, the presence of underlying lattice enables anyonic excitations in the FQAH states to exhibit a non-negligible and non-trivial dispersion. When the band filling factor is doped away from FQAH insulators, novel superconducting states can emerge (*68-70*). The proposed superconductor pairing mechanism is unconventional, which could coexist with spontaneous time reversal symmetry (TRS) breaking. Our experimental results reveal an intricate interplay among competing superconducting phases, RIQAH states, and FQAH states within a single moiré Chern band, suggesting an intrinsic connection between the superconducting pairing mechanism and emergent anyonic excitations. Further experiments, such as optical Kerr rotation, magnetic circular dichroism spectroscopy, and scanning SQUID measurements, are needed to confirm the TRS breaking nature of the observed superconductivity.

In conclusion, we have observed a series of IQAH, FQAH, RIQAH states, and signatures of unconventional superconductivity in the first moiré Chern band of ~3.83°-3.93° $tMoTe_2$. Our findings not only open up an exciting possibility for exploring exotic superconductivity and anyonic quantum matters beyond conventional Landau-level paradigm within moiré Chern bands, but also pave the way for interfacing IQAH and FQAH states with superconductivity based on $tMoTe_2$ by simple electrostatic gating. This could be highly desirable for engineering Majorana or parafermion zero modes towards topological quantum computation.

**Acknowledgement**

We sincerely thank Rui-Rui Du, Zhao Liu, Xiaoyan Xu, and Xiaoxiang Xi for helpful discussions. This work is supported by the National Key R&D Program of China (Nos. 2022YFA1405400, 2022YFA1402702, 2022YFA1402404, 2022YFA1403301,

2021YFA1401400, 2021YFA1400100, 2021YFA1202902), the National Natural Science Foundation of China (Nos. 12350403, 92565302, 12374045, 12350404, 12204115, 12374292, 12488101, 12461160252, T2525009, 92265102), the Quantum Science and Technology-National Science and Technology Major Project (Nos. 2021ZD0302600, 2021ZD0302500, 2024ZD0300104), the Science and Technology Commission of Shanghai Municipality (Grants Nos. 24QA2703700, 24LZ1401100, 23JC1400600, 2019SHZDZX01). T.L., X.L. and S.J. acknowledge the Shanghai Jiao Tong University 2030 Initiative Program. T.L. and Yuanbo Zhang acknowledges support from New Cornerstone Science Foundation through the New Cornerstone Investigator Program and the XPLORER PRIZE. T.L. acknowledges support from the Asian Young Scientist Fellowship. S.L. acknowledges support by the Hong Kong RGC (No. 26308524, AoE/P-604/25R), the Hong Kong Collaborative Research Fund (No. C6053-23G), the Guangdong Provincial Quantum Science Strategic Initiative (GDZX2501003), and the Ministry of Science and Technology of the People's Republic of China (No. MOST23SC01). C.X. and Yang Zhang acknowledge support from the Max Planck partner lab grant for quantum materials. J.L. acknowledges Quantum Science Strategic Special Project (No. GDZX2301006), and Shenzhen Municipal Funding Co-construction Program Project (No. SZZX2301004). K.W. and T.T. acknowledge support from the JSPS KAKENHI (Nos. 21H05233 and 23H02052) and World Premier International Research Center Initiative (WPI), MEXT, Japan. A portion of this work was carried out at the Synergetic Extreme Condition User Facility (SECUF, https://cstr.cn/31123.02.SECUF).

**Competing interests**

The authors declare no competing financial interests.

## Figures

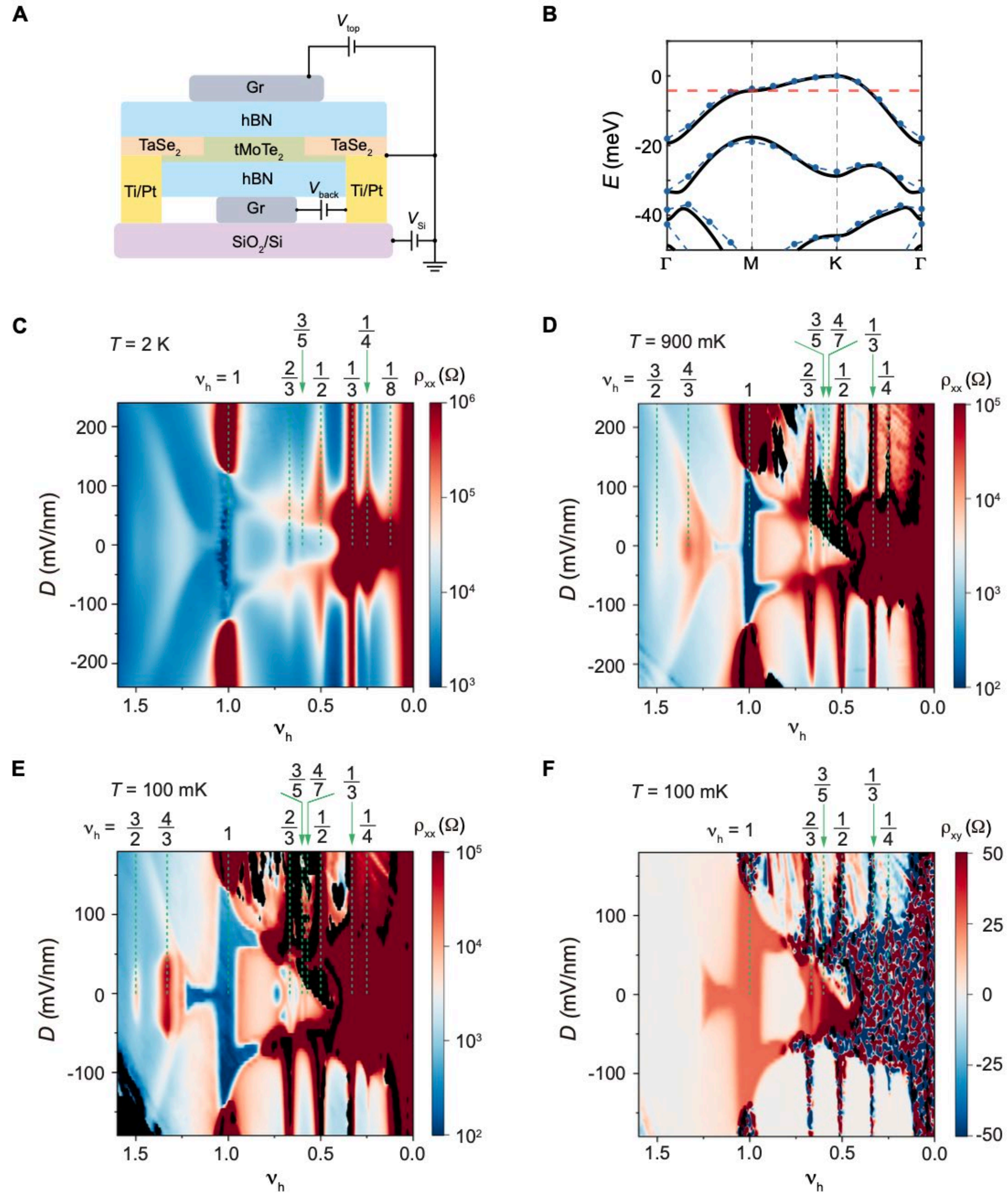

**Fig. 1 | Transport phase diagram of high-quality $tMoTe_2$ (Device 1).** (**A**) Schematic of the triple-gated $tMoTe_2$ device used for transport measurements. (**B**) Calculated band structure of 3.89° $tMoTe_2$, where the black line represents the results from the continuum model, the blue dots correspond to the DFT bands and the red dashed line indicates the energy of the van Hove singularity. (**C** and **D**) Longitudinal resistivity $\rho_{xx}$ as a function of electric displacement field $D$ and moiré hole filling factor $\nu_h$ at $T = 2$ K (C) and $T = 900$ mK (D), measured at $B = 0$. (**E** and **F**) Symmetrized $\rho_{xx}$ (E) and anti-symmetrized $\rho_{xy}$ (F) under $B = \pm 0.1$ T, as a function of $D$ and $\nu_h$ at $T = 100$ mK. The black regions in (D-F) are experimentally inaccessible, either due to their highly insulating nature or contact issues. Dashed lines in (C-F) indicate the correlated and/or topological quantum states that emerge at commensurate moiré fillings.

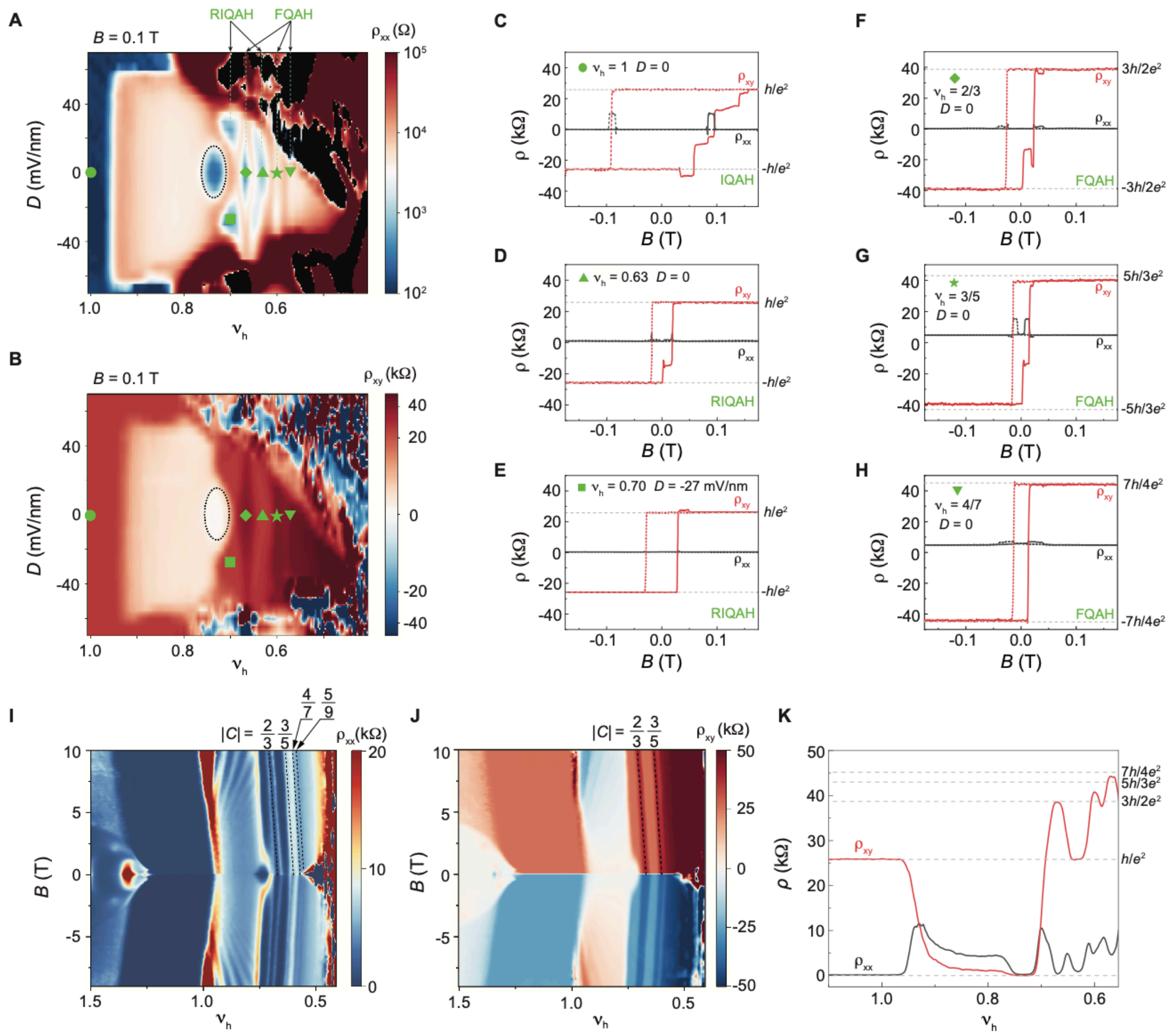

**Fig. 2 | IQAH effect, FQAH effect and RIQAH effect in 3.83° $tMoTe_2$ (Device 1).** (**A** and **B**) Symmetrized $\rho_{xx}$ (A) and anti-symmetrized $\rho_{xy}$ (B) under $B = \pm0.1$ T, as a function of $D$ and $\nu_h$. FQAH and RIQAH states are indicated by green dashed lines in (A). The black dashed circle depicts the superconducting region with both vanishing $\rho_{xx}$ and vanishing $\rho_{xy}$. (**C-H**) $B$-dependent $\rho_{xx}$ and $\rho_{xy}$ at the $\nu_h$, $D$ points highlighted by green markers shown in (A) and (B). Solid (dashed) lines represent scans of $B$ from negative (positive) values to positive (negative) values. $\rho_{xx}$ in (C-H) has been symmetrized with respect to $B$, while $\rho_{xy}$ is shown after subtracting an offset from the raw data, without being anti-symmetrized. (**I** and **J**) Raw data of $\rho_{xx}$ (I) and $\rho_{xy}$ (J) as a function of $B$ and $\nu_h$ at $D = 0$. Dashed lines in **(I-J)** represent the expected dispersions based on Streda formula for the FQAH states at $\nu_h$ = 2/3, 3/5, 4/7, and 5/9 with $|C|$ = 2/3, 3/5, 4/7, and 5/9, respectively. **(K)** Symmetrized $\rho_{xx}$ and anti-symmetrized $\rho_{xy}$ under $B = \pm0.1$ T, as a function of $\nu_h$ at $D = 0$. Quantized or nearly quantized plateaus of $\rho_{xy}$, and local minima of $\rho_{xx}$ can be clearly seen for the IQAH, FQAH, and RIQAH states. Grey dashed lines represent zero resistance, and the expected quantized values for the IQAH, FQAH, and RIQAH states. The measurement temperature for (A-K) is $T_{MC}$ = 15 mK.

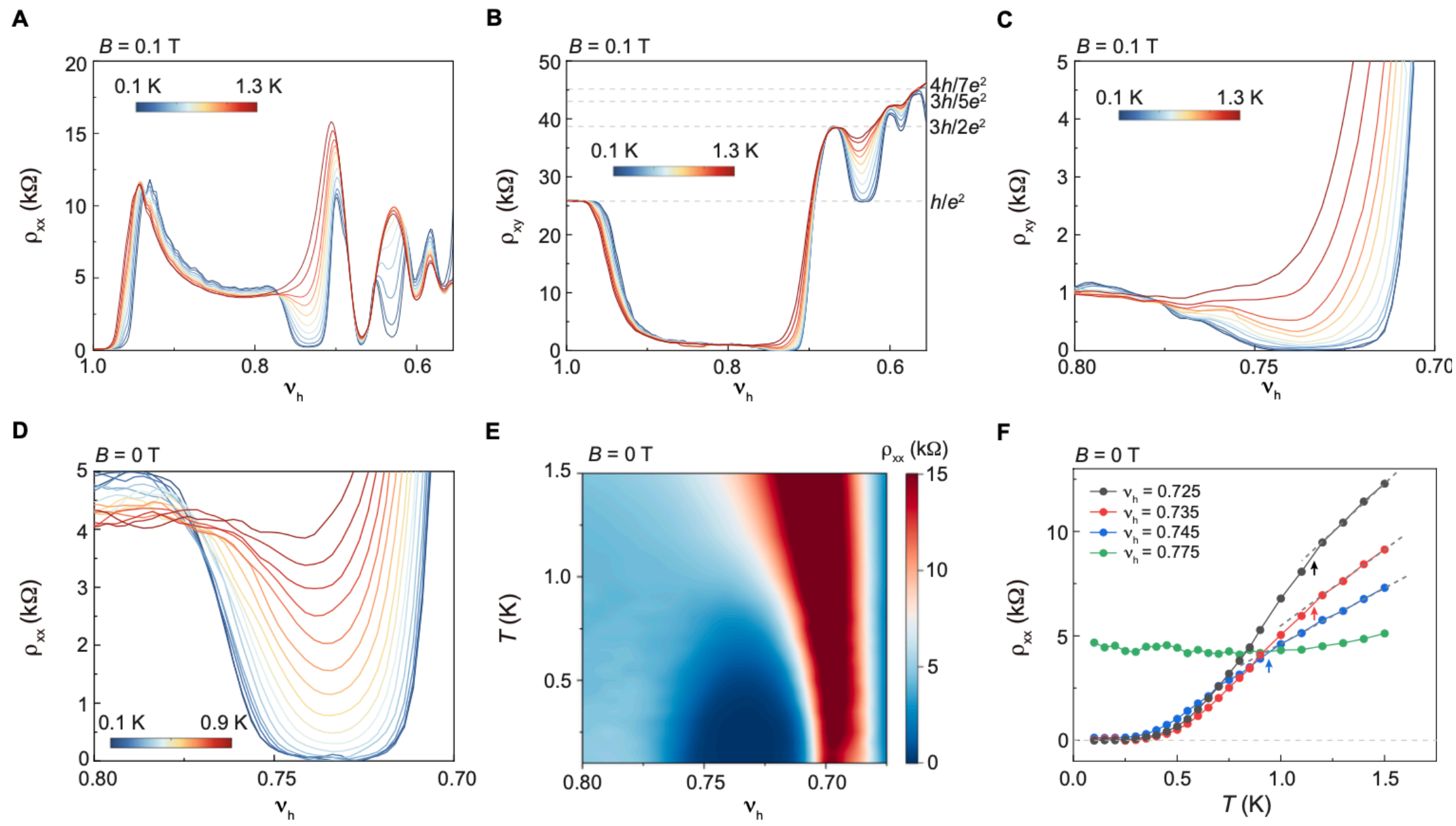

**Fig. 3 | Quantized anomalous Hall plateaus and superconductivity in 3.83° $tMoTe_2$ (Device 1).** (**A** and **B**) Temperature dependence of symmetrized $\rho_{xx}$ (A) and anti-symmetrized $\rho_{xy}$ (B) versus $\nu_h$ under $B = \pm 0.1$ T at $D = 0$. Both the RIQAH states and the superconducting state have smaller energy scale than the FQAH states. (**C**) The zoomed-in view of (B), highlighting that $\rho_{xy}$ approaches zero in the superconducting state at low temperatures. The temperature points in (A-C) include 100 mK to 900 mK in steps of 100 mK, and 1.1 K, 1.3 K. (**D**) Temperature dependence of $\rho_{xx}$ versus $\nu_h$ at $B = 0$ and $D = 0$. The temperature points range from 100 mK to 900 mK, in 50 mK steps. (**E**) $\rho_{xx}$ map as a function of $T$ and $\nu_h$ at $B = 0$ and $D = 0$, showing the superconducting dome. (**F**) Linecuts of **e** at representative $\nu_h$. The onset transition temperature of superconductivity $T_{onset}$ is indicated by arrows, where the measured $\rho_{xx}$ deviates from the projected normal-state resistivity (dashed lines).

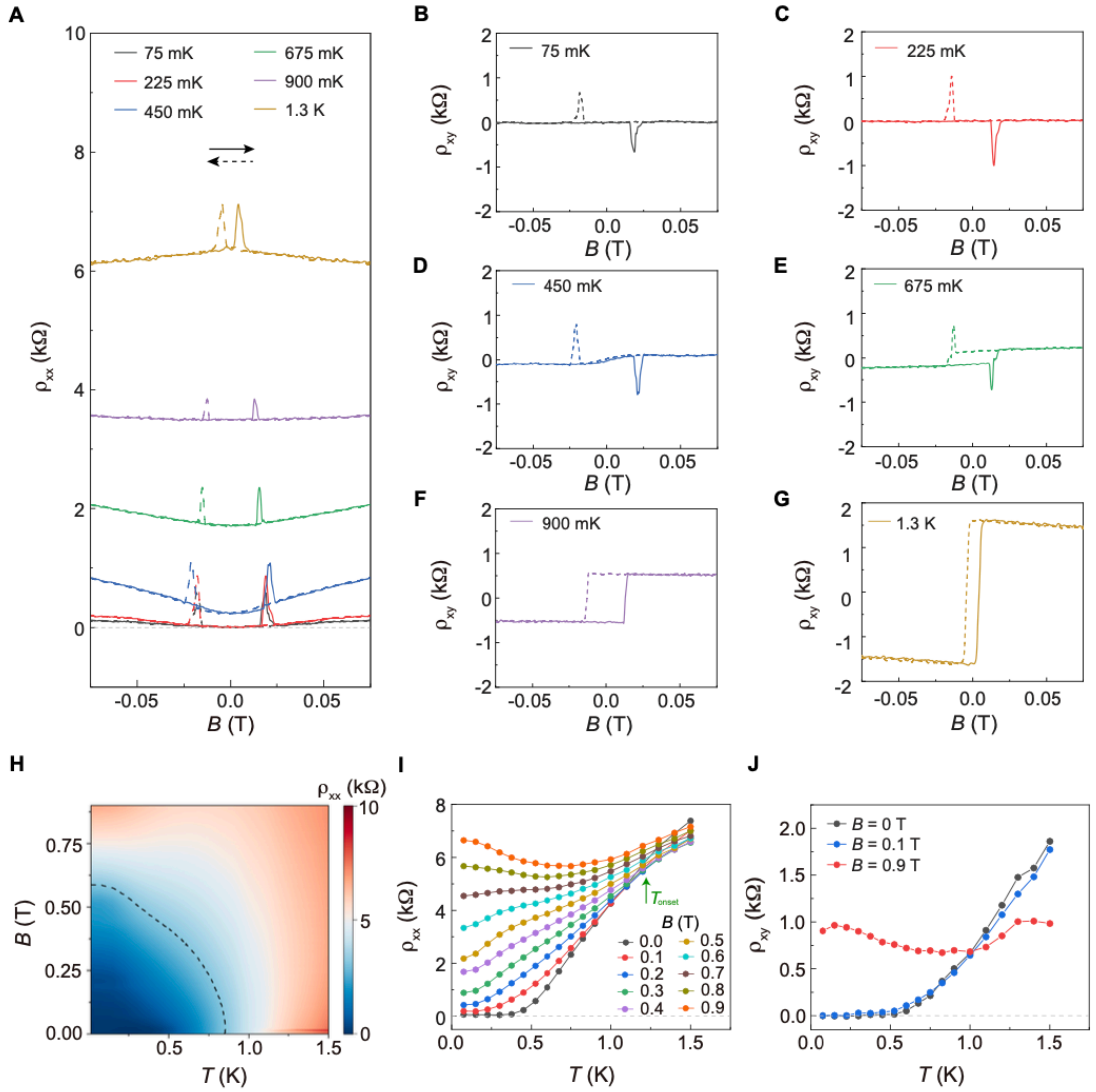

**Fig. 4 | Temperature and perpendicular magnetic field dependence of the superconducting state. (A)** $B$-dependent $\rho_{xx}$ in the superconducting state at $T$ = 75 mK, 225 mK, 450 mK, 675 mK, 900 mK, and 1.3 K. (**B-G**), $B$-dependent $\rho_{xy}$ in the superconducting state, measured at the same temperature points as in (A). In (A-G), Solid (dashed) lines represent scans of $B$ from negative (positive) values to positive (negative) values. (**H**) Map of $\rho_{xx}$ as a function of $T$ and $B$, with the superconducting critical magnetic field $B_c$ marked by the dashed line. (**I**) Linecuts of **e** at representative $B$. $T_{onset}$ is marked by the arrow. (**J**) $\rho_{xy}$ as a function of $T$ at $B$ = 0 T, 0.1 T and 0.9 T. Dashed lines in (H and J**)** represent zero resistance. All data in this figure are measured at $\nu_h$ = 0.74 and $D$ = 0 in Device 1.

## Materials and Methods

2H-$MoTe_2$ crystal growth

High-quality single crystals of 2H-$MoTe_2$ were grown via an optimized vertical self-flux approach. A mixture of molybdenum powder (99.997% purity) and tellurium chunks (99.9999% purity) with a molar ratio of 1:110 was placed in a quartz ampoule. Following several purge-pump cycles, the ampoule was sealed under a vacuum of $\sim 10^{-5}$ Torr. The growth process involved heating the ampoule to 880 °C over a 24-hour period, keeping it isothermal for 12 hours, and then cooling it at a rate of 1 °C/h down to 530 °C. At this temperature, the excess Te flux was removed by centrifugation. To extract any residual Te and minimize inclusions, a final post-annealing treatment was conducted for 48 hours in a thermal gradient ($T_{hot}$ = 400 °C, $\Delta T \sim$ 200 °C). The resulting $MoTe_2$ crystals are highly suited for mechanical exfoliation and 2D device fabrication.

Device fabrications

The t$MoTe_2$ devices are fabricated with a triple-gated geometry, as described in Ref. 4 and 9. Atomic thin flakes of graphite, h-BN, 2H-$MoTe_2$, and 2H-$TaSe_2$ are obtained through mechanical exfoliation from bulk crystals. Flux-grown 2H-$MoTe_2$ crystals and commercial (HQ-graphene) 2H-$TaSe_2$ are used for exfoliation. The stack is made by the standard dry transfer method (*71*) with a polycarbonate stamps inside a nitrogen-filled glove box. The stacking sequence, from top to bottom, is as follows: graphite as the top gate electrode, h-BN (10-15 nm) as the top dielectric layer, few-layer $TaSe_2$ as the metallic electrodes, t$MoTe_2$ (in contact with the $TaSe_2$ layer, with the twisted angle controlled by a mechanical rotator), h-BN (10-15 nm) as the bottom dielectric layer, and graphite as the bottom gate electrode. The entire stack is released onto a Si/$SiO_2$ substrate, with prepatterned Ti/Pt (3 nm/10 nm) electrodes. The finished stack was subsequently annealed at 300 °C for approximately 1 hour in Ar/$H_2$ forming gas. We found that this annealing process helps improve both the contact yield and the moiré uniformity. The stack is processed into a standard Hall bar device using standard e-beam lithography and reactive ion etching techniques.

Among the devices we fabricated, we first perform measurements at 1.6 K to obtain a basic characterization of sample quality and contact resistance. About one-third of the devices allow for precise twist-angle calibration via quantum oscillations, while the remaining two-thirds do not exhibit observable oscillations—likely due to limitations in moiré uniformity or contact quality—and therefore their twist angles cannot be accurately determined. Within the 3.75°– 4° range, we have 25 samples with calibrated twist angles. From these, we selected seven high-quality devices for further measurements in dilution refrigerators at ultra-low temperatures, among which three exhibit superconductivity near the FQAH and RIQAH states (Devices 1-3).

Transport measurements

Electrical transport measurements were performed using $^4$He cryostats (PPMS, Oxford Teslatron, with base temperatures of ~ 1.5 -2 K) and dilution refrigerators. All devices were first characterized at $^4$He cryostats and subsequently reloaded into dilution refrigerators for ultra-low temperature measurements. In particular, Device 1 was first measured in a top-loading dilution refrigerator (Oxford TLM) equipped with an 18 T superconducting magnet. The sample was immersed in the $^3$He-$^4$He mixtures during measurements. The nominal base temperature of the $^3$He-$^4$He mixtures is about 15 mK, and the accessible temperature range is 15 mK to 1.5 K. Each fridge line has a sliver epoxy filter and a low-temperature RC- filter (consisting of a 470 Ω resistor and a 100-pF capacitor). The base electron temperature for graphene devices is estimated to be around 100 mK, based on prior experience with measuring superconductivity in Bernal bilayer graphene (*36*). Device 1 was also measured later in a cryogen free dilution refrigerator (Q-one, Q-400) equipped with a 9 T superconducting magnet. Device 2-6 were measured in the same system. Each measurement line in this dilution refrigerator was equipped with a sliver epoxy filter and multistage RC- filters (consisting of ~1500 Ω resistors and ~3 nF capacitors in total, mounted on the mixing chamber plate). The nominal base temperature of mixing chamber is 10 mK.

We performed the electrical transport measurements by using the standard low-frequency lock-in techniques. Electrical contacts to $tMoTe_2$ are achieved by using 2D metal $2H\text{-}TaSe_2$ together with $Si/SiO_2$ gate to induce heavily hole-doping in contact regions. The bias current is limited within 1 nA to avoid sample heating and disturbance of fragile quantum states. We convert the measured longitudinal resistance $R_{xx}$ into longitudinal resistivity $\rho_{xx}$ by $\rho_{xx} = R_{xx}\frac{W}{L}$, where $W$ is the Hall bar width and $L$ is the separation between voltage probes. The Hall resistivity $\rho_{xy}$ equals to measured Hall resistance $R_{xy}$ in two-dimensional case.

Calibration of moiré filling factors

The device geometry allows us to independently tune the carrier density $\left(n = \frac{c_t V_t + c_b V_b}{e} + n_0\right)$ and the vertical electric displacement field $\left(D = \frac{c_t V_t - c_b V_b}{2\varepsilon_0} + D_0\right)$ in $tMoTe_2$ by applying top graphite gate voltage $V_t$ and bottom graphite gate voltage $V_b$. Here, $\varepsilon_0$, $c_t$, $c_b$, $n_0$ and $D_0$ denote the vacuum permittivity, geometric capacitance of the top graphite gate, geometric capacitance of the bottom graphite gate, intrinsic doping and the built-in electric field, respectively. The value of $c_t$ and $c_b$ are mainly determined by measuring Shubnikov–de Haas (SdH) oscillations in $R_{xx}$ under perpendicular magnetic field $B$ (fig. S2), and double checked by measuring the thickness of *h*-BN layers. We convert $n$ to moiré filling factor $\nu$ using the density difference between a series of correlated and/or topological quantum states with prominent $\rho_{xx}$ and $\rho_{xy}$ features. The hole density corresponding to $\nu_h = 1$ is defined as the moiré density $n_M$, representing the carrier density required to fill one hole per moiré unit cell. The twist angle $\theta$ and moiré wavelength $a_M$ are then determined from $n_M = \frac{2}{\sqrt{3}a_M^2}$ and $a_M = \frac{a}{\sqrt{2(1-\cos\theta)}}$, where $a$ =0.352 nm is the lattice constant of monolayer $MoTe_2$ (*72*). The values of $\theta$, $a_M$, $n_M$ for device 1-6 are summarized in the Table 1.

More about the superconducting state

In the two-dimensional limit, due to enhanced thermal and quantum fluctuations, a true superconducting state with quasi-long-range order exists only below the Berezinskii-Kosterlitz-Thouless (BKT) transition temperature $T_{BKT}$. Although electron pairing initiates at higher temperatures (characterized by $T_{onset}$), electronic transport between $T_{BKT}$ and $T_{onset}$ is dissipative due to the unbinding vortices, resulting in finite resistance. Moreover, an anomalous metallic (or "failed superconductor") phase has been observed in 2D superconducting systems (*73*, *74*), typically under a small magnetic field but still well below the $B_c$. This phase is characterized by a nonzero $\rho_{xx}$ and vanishing $\rho_{xy}$, as what we observed and shown in Fig. 4A-4G.

The Ginzburg-Landau superconducting coherence length $\xi$ can be estimated based on the $B_c$ measured at $T \ll T_c$ from the relation $\xi = \sqrt{\Phi_0/(2\pi B_c)}$. Here $\Phi_0 = h/2e$ is the superconducting flux quantum. Based on the data shown in Fig. 4H and Fig S8, the $B_c$ at base temperature is approximately 0.6-0.8 T, resulting in a $\xi \approx$ 20-23 nm. The averaging distance between holes at $\nu_h$ = 0.73 for 3.83° tMoTe$_2$ is $d_{hole} \approx 1/\sqrt{\nu_h n_M} \approx$ 6 nm, leading to $\frac{\xi}{d_{hole}} \approx$ 3.3-3.8, suggesting a strong coupling of Cooper pairing.

Starting from a two-dimensional Drude model and assuming a parabolic band dispersion, the mean free path $l_m$ can be estimated as $l_m = \frac{h}{e^2 k_F \rho_{xx}}$ , where $\rho_{xx}$ is normal-state resistivity in the superconducting regime (~ 5 kΩ for our case) and $k_F = \sqrt{4\pi n}$ is the Fermi wave vector, assuming a degeneracy of 1. This estimation yields a $l_m \approx$ 17 nm, if we take the carrier density $n$ to correspond to the moiré filling associated with the superconducting state (i.e., counting from the band edge). This value is comparable to the estimated superconducting coherence length. However, for moiré flat bands, the exact band dispersion is unknown and can be strongly renormalized by correlation effects, leading to significant deviations from a simple parabolic form. Moreover, here the carrier density $n$ cannot be accurately determined from Hall measurements because the superconducting regime is close to a VHS-like feature and is complicated by the presence of an anomalous Hall signal. These factors make an accurate estimate of the mean free path challenging. On the other hand, the relatively broad superconducting transition we observe may suggest that the system is not fully in the clean limit, and that the observed exotic superconducting states are sensitive to disorders. Further investigations using even higher-quality samples will be necessary to clarify this.

Band structure calculations

As described in Ref. *75*, we begin by constructing a neural network force field model to accurately relax the structure of twisted MoTe$_2$. Next, we compute the band structure using the Vienna Ab initio Simulation Package (VASP). To ensure precise agreement with the bands calculated by density functional theory (DFT) methods, we extend the continuum model by including the second harmonics terms for both inter-layer and intra-layer coupling. With fitting method, we arrive the parameter:

$$m^* = 0.62m_0,\ V_1 = 10.3\ \text{meV},\ V_2 = 2.9\ \text{meV},\ w_1 = -7.8\ \text{meV},\ w_2 = 6.9\ \text{meV},\ \varphi = -75°$$

Here, $m^*$ represents the effective mass of MoTe$_2$, $m_0$ is the mass of free electron, $V_1$ and $V_2$ denote the strengths of the first and second harmonic terms of the intra-layer potential, $\varphi$ is the phase of the first harmonic term of $V_1$, and $w_1$, $w_2$ correspond to the strengths of the first and second harmonic terms of the inter-layer tunneling. This refined model effectively reproduces all DFT bands within a twist angle range of 3° to 5°. Finally, we diagonalize the continuum model Hamiltonian and compute the density of states (DOS) using the following formula:

$$DOS(E) = \frac{1}{\mathrm{c}\sqrt{2\pi}\Omega}\sum_{\mathrm{nk}} e^{-\frac{(E_{\mathrm{nk}}-E)^2}{2\mathrm{c}^2}}$$

where $E_{\mathrm{nk}}$ represents the energy spectrum, $\Omega$ is the area of the moiré unit cell and $c$ (set to be 1 meV here) is the broadening parameter that controls the width of the Gaussian broadening applied to the DOS.

**Table I. Device summary table**

| Device | Max monolayer Hall mobility ($cm^2$/Vs) * | $\theta$ (°) ** | $a_M$ (nm) | $n_M$ ($10^{12}$ $cm^{-2}$) | The Onset *B* of SdH oscillations at *D* = 0 *** | RIQAH | SC near FQAH |
|---|---|---|---|---|---|---|---|
| D1 | ~ 10000 | 3.83±0.02 | 5.27±0.03 | 4.16±0.05 | ~ 3 T | Y | Y |
| D2 | ~ 29000 | 3.89±0.03 | 5.19±0.04 | 4.29±0.07 | ~ 5 T | Y | Y |
| D3 | ~ 40000 | 3.93±0.04 | 5.14±0.05 | 4.37±0.09 | ~ 6 T | Y | Y |
| D4 | ~ 40000 | 3.79±0.03 | 5.32±0.04 | 4.07±0.08 | ~ 5 T | Y | N |
| D5 | ~ 10000 | 3.98±0.03 | 5.07±0.04 | 4.49±0.07 | ~ 5 T | Y | N |
| D6 | ~ 15000 | 3.90±0.06 | 5.18±0.08 | 4.31±0.13 | ~ 9 T | Y | N |

*The 'Max monolayer Hall mobility' refers to the highest Hall mobility value observed at *T* ~ 1.5 K -2 K in monolayer $MoTe_2$ devices fabricated from the same batch of bulk crystal as the moiré devices. In total, four batches of flux-grown 2H-$MoTe_2$ crystals were used, with the maximum monolayer Hall mobility ranging from 10,000 $cm^2$/Vs to 40,000 $cm^2$/Vs. Typical transport characteristics of a high-quality monolayer $MoTe_2$ device are shown in Fig. S1.

**The error bars in $\theta$, $a_M$, and $n_M$ include not only the fitting uncertainty, but also the variation in the extracted values between neighboring contact pairs (see Fig. S2 as an example). These error bars serve as a useful indicator for the moiré inhomogeneity.

***The onset magnetic field of the Shubnikov–de Haas (SdH) oscillations between $\nu_h$ = 2/3 and 1 at *D* = 0 can serve as an empirical indicator of the overall device quality. Resolving SdH oscillations in this regime depends on both the single-crystal quality and the moiré quality, and such oscillations were not resolved in previous works (*3*, *4*, *9*, *10*).

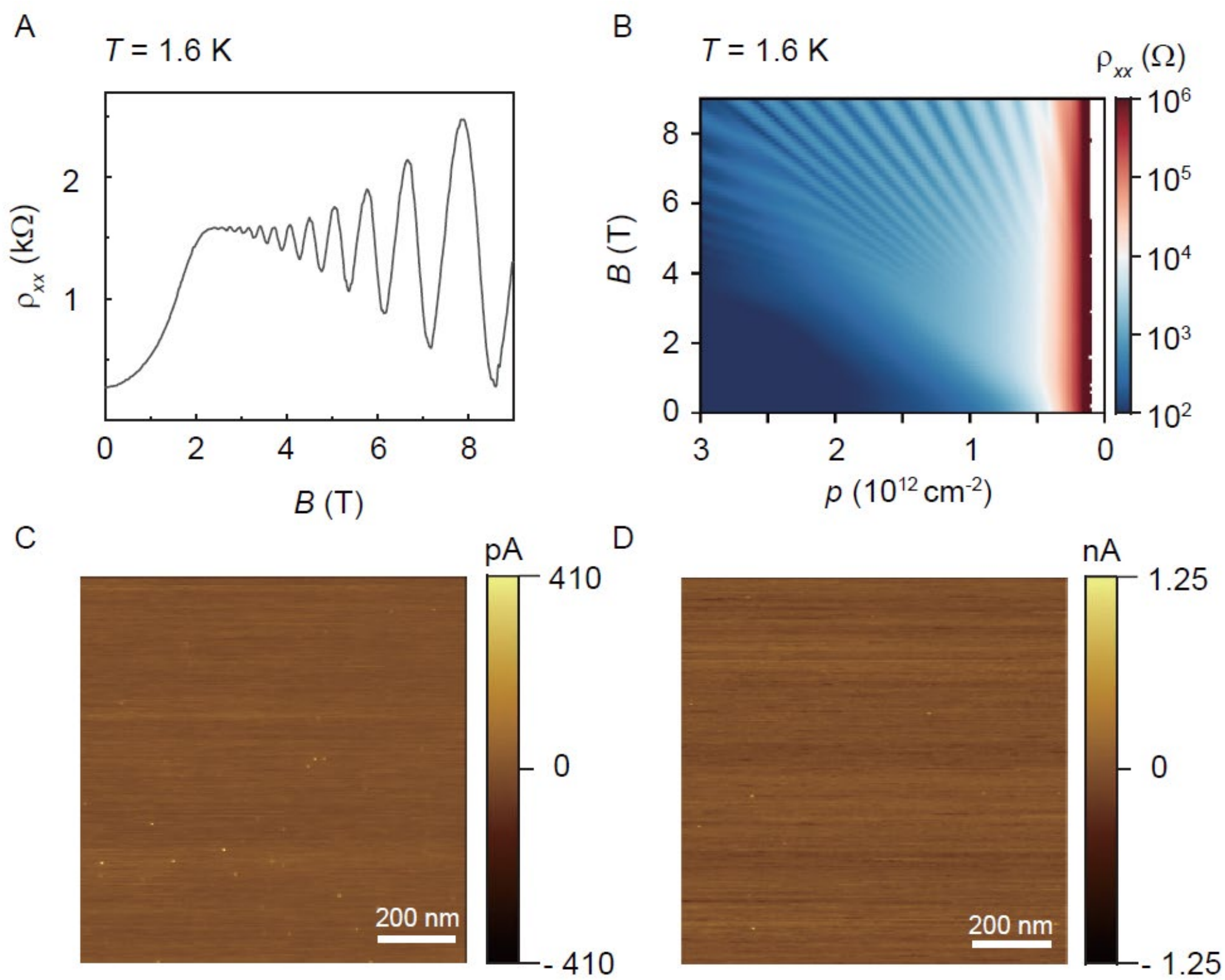

**Fig. S1. Characterizations of flux-grown, high-quality 2H-MoTe$_2$ crystals.** (**A**) Longitudinal resistivity $\rho_{xx}$ as a function of $B$ of a monolayer MoTe$_2$ device made by flux-grown crystal, measured at $T$ = 1.6 K and hole doping density $p \approx 1\times10^{12}$ cm$^{-2}$. The device structure is similar to that used for the tMoTe$_2$ devices (Fig. 1A), where electrical contact to the monolayer MoTe$_2$ is achieved using TaSe$_2$ and Si/SiO$_2$ gating. Well resolved Shubnikov–de Haas (SdH) oscillations can be observed when $B$ above approximately 2 T. The same batch of bulk crystals was used to fabricate twisted MoTe$_2$ device D2. (**B**) $\rho_{xx}$ as functions of $B$ and hole density at $T$ = 1.6 K for the monolayer MoTe$_2$ device. The maximum Hall mobility for this particular device is about 29000 cm$^2$/Vs at 1.6 K. (**C-D**) Conductive atomic force microscopy (c-AFM) scans of exfoliated bulk flux-grown MoTe$_2$ crystals. The scan area is 1 μm×1 μm, and a bias voltage of 1.0 V (C) and 0.85 V (D) was applied to the AFM tip during the measurements, respectively. The scans were performed inside a nitrogen-filled glovebox with oxygen and moisture level below 0.1 ppm. In transition metal dichalcogenides, defects are generally classified into two main categories: charged defects and isovalent defects (*76*). Large-area c-AFM scans are primarily sensitive to charged defects (*77*). The defect densities resolved by c-AFM are approximately $3\times10^9$ cm$^{-2}$ for (C) and $1\times10^9$ cm$^{-2}$ for (D), respectively. The crystal shown in the left panel (C) is from the same batch used to fabricate the monolayer device shown in (A–B) as well as the tMoTe$_2$ device D2. The crystal shown in the right panel (D) is from the same batch used to fabricate tMoTe$_2$ devices D3 and D4.

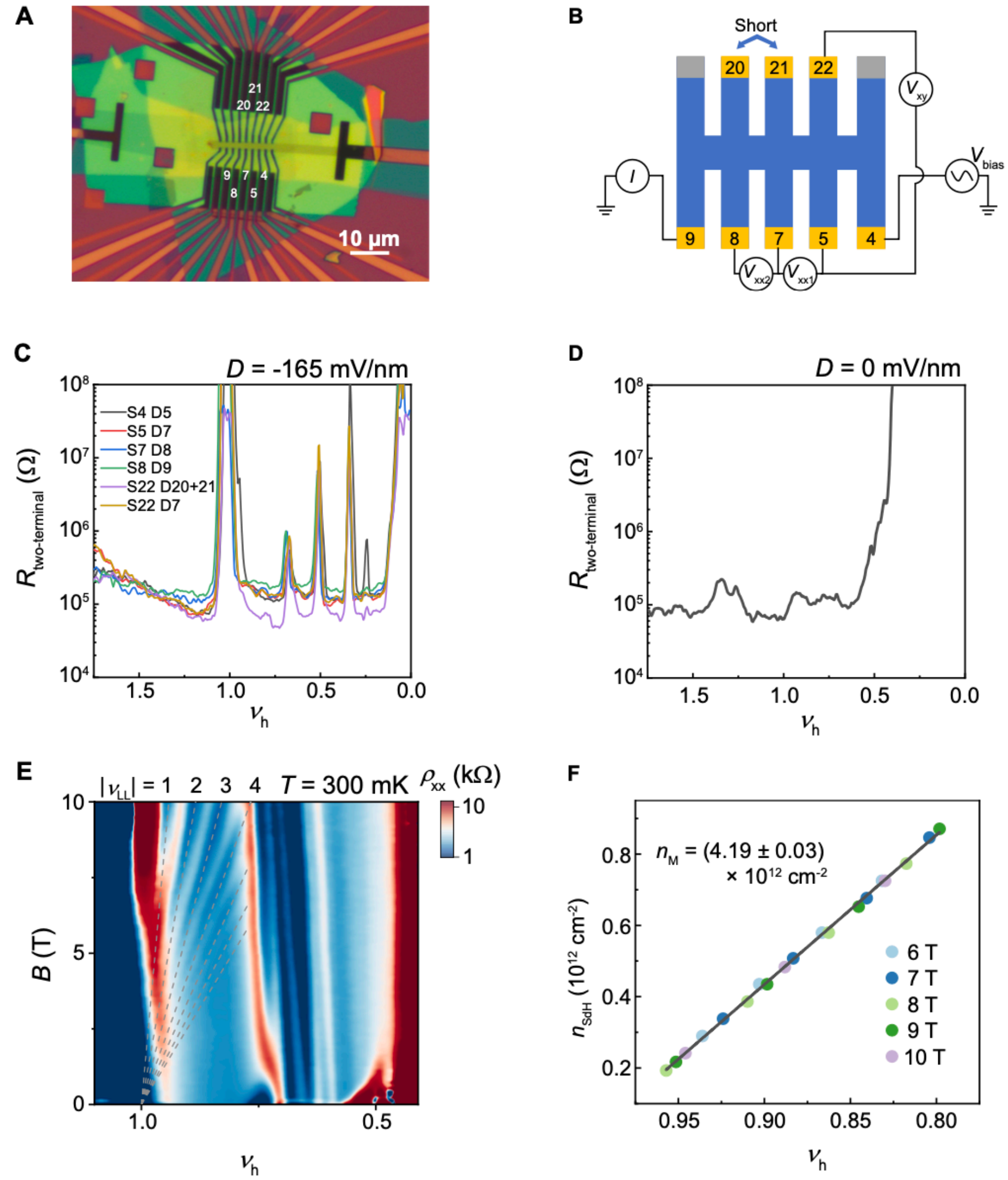

**Fig. S2. Device geometry, twist angle calibration and sample inhomogeneity of Device 1.** (**A**) Optical microscope image of Device 1 (3.83° $tMoTe_2$). The scale bar is 10 μm. (**B**) Schematic of the Hall bar geometry and a typical measurement configuration. Functional contacts are labeled by numbers in (A-B), where contacts 20 and 21 are shorted. (**C**) Two-terminal resistance $R_{\text{two-terminal}}$ as a function of $\nu_h$ at $D$ = -165 mV/nm and $T_{MC}$ = 15 mK, measured with different configurations. Features of correlated insulators measured with different configurations align well, demonstrating the high uniformity of the device. The variation in $\nu_h$ across the entire Hall bar is within 0.03. (**D**) $R_{\text{two-terminal}}$ as a function of $\nu_h$ at $D$ = 0 and $T_{MC}$ = 15 mK, measured with contacts 5 and 7. The contact resistance is estimated about half of the $R_{\text{two-terminal}}$, which is about 50 kΩ for $\nu_h > 0.55$. (**E**) Symmetrized $\rho_{xx}$ as a function of $\nu_h$ and $B$ measured with voltage probe 7 and 8, as shown in (B), at $D$ = -5 mV/nm and $T$ = 300 mK. A set of Landau fan emanates from $\nu_h$ = 1 are highlighted by dashed lines, and the Landau level filling factors $\nu_{LL}$ from 1 to 4 are labeled correspondingly. The onset of SdH oscillations occurs around 3 T, and they are single-fold degenerate. (**F**) At several fixed $B$, we calculate the carrier density based on SdH oscillations using $n_{SdH} = \nu_{LL}eB/h$, and plot them as a function of $\nu_h$. From the slope of the linear fit, we obtain a moiré density $n_M = (4.19 \pm 0.03) \times 10^{12}\ \text{cm}^{-2}$. Similarly, using the data measured with voltage probe 5 and 7, we get $n_M = (4.13 \pm 0.03) \times 10^{12}\ \text{cm}^{-2}$. Considering $n_M$ values from both measurement configurations, the twist angle is determined to be 3.83°±0.02°.

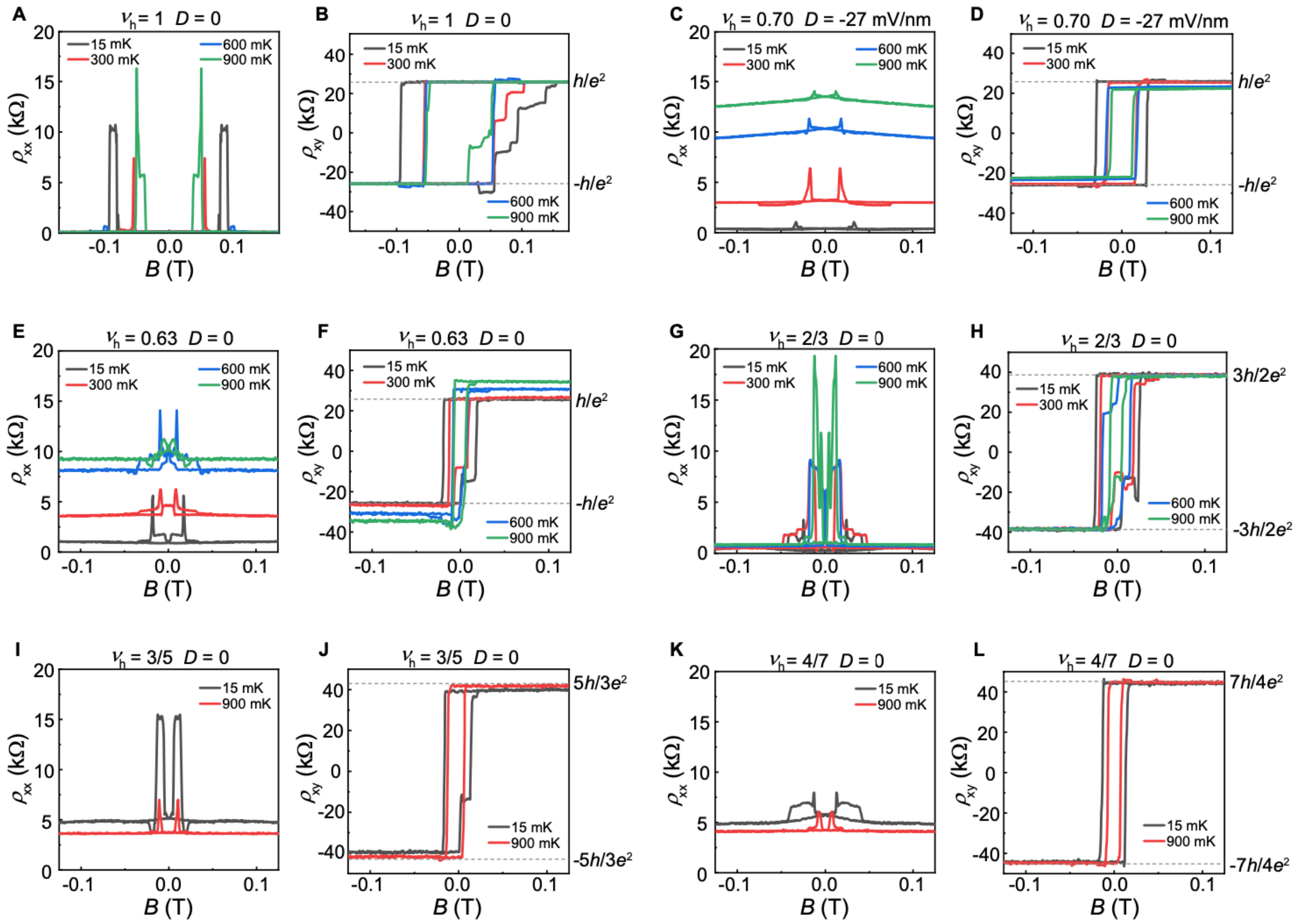

**Fig. S3. Temperature dependence of magnetic hysteresis scans for the IQAH, FQAH, and reentrant IQAH states.** (**A-L**) $B$-dependent $\rho_{xx}$ and $\rho_{xy}$ of Device 1 measured at varying temperatures of the IQAH state (A-B), the RIQAH states (C-F), and the FQAH states (G-L). In these plots, $\rho_{xx}$ has been symmetrized with respect to $B$, while $\rho_{xy}$ is shown after subtracting an offset from the raw data, without being anti-symmetrized. The specific $\nu_h$, $D$ values are labeled in each figure correspondingly.

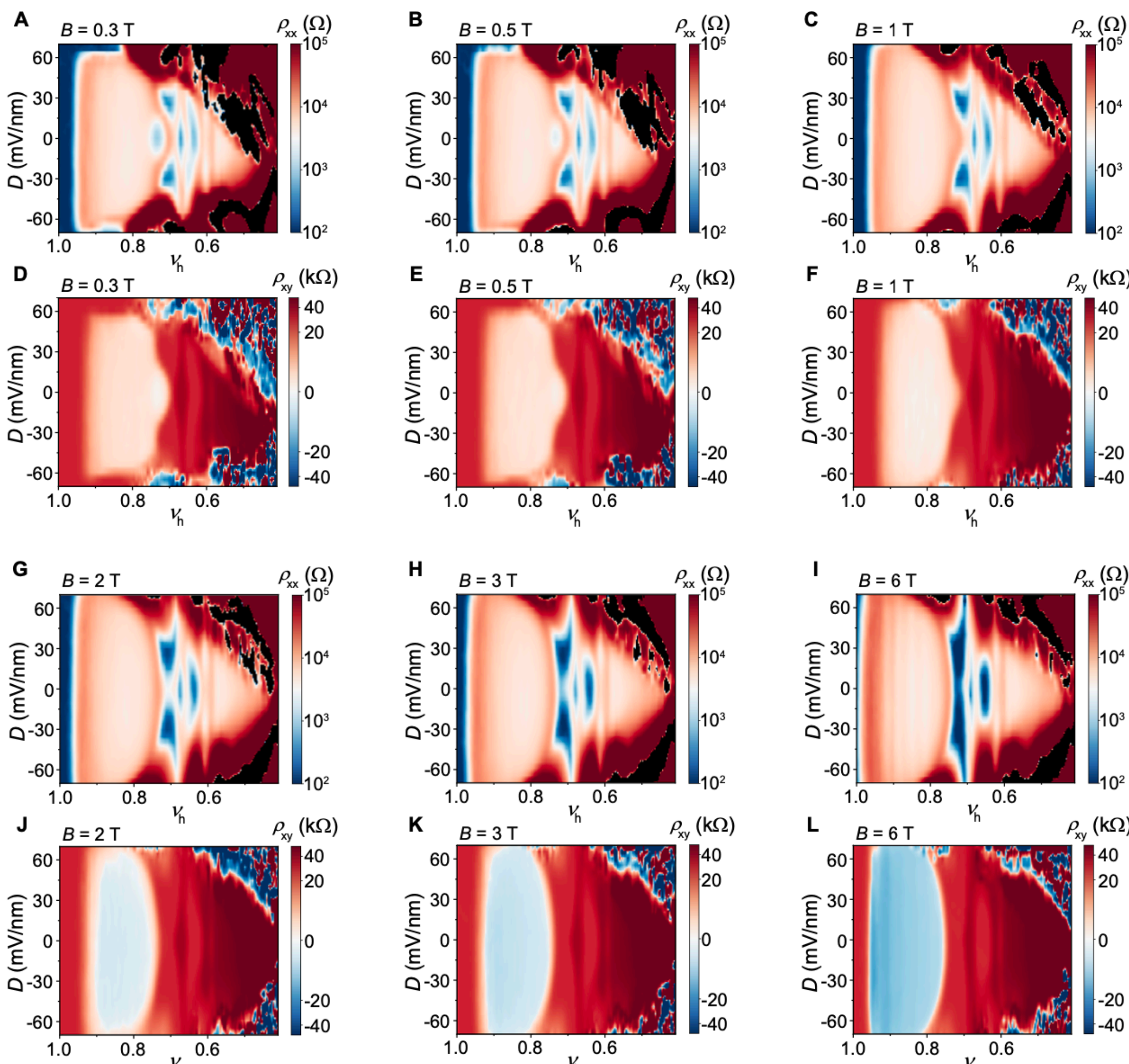

**Fig. S4. Perpendicular magnetic field dependence of $\rho_{xx}$-$\nu_h$-$D$ and $\rho_{xy}$-$\nu_h$-$D$ maps at base temperature.** (**A-L**) Symmetrized $\rho_{xx}$ and anti-symmetrized $\rho_{xy}$ as a function of $\nu_h$ and $D$, measured at varying $B$ (0.3 T, 0.5 T, 1 T, 2 T, 3 T, 6 T) and $T_{MC}$ = 15 mK in Device 1. As the magnetic field increases, the superconducting state is gradually suppressed. In contrast, the $\nu_h \sim 0.7$ RIQAH state extends over a broader $D$-field range with increasing $B$, eventually taking over the superconducting region at sufficiently high $B$-field. The existence of VHS features is evident by the sign change in $\rho_{xy}$ near $\nu_h \approx 0.7$, accompanied with a local peak in $\rho_{xx}$ at $D = 0$, which shifts towards higher values of $\nu_h$ as $B$ and/or $D$ increase. Note that the $\rho_{xy}$ sign change can only be observed at $B \geq 2$ T, presumably due to the anomalous Hall signals dominant over the normal Hall signals at low $B$-fields.

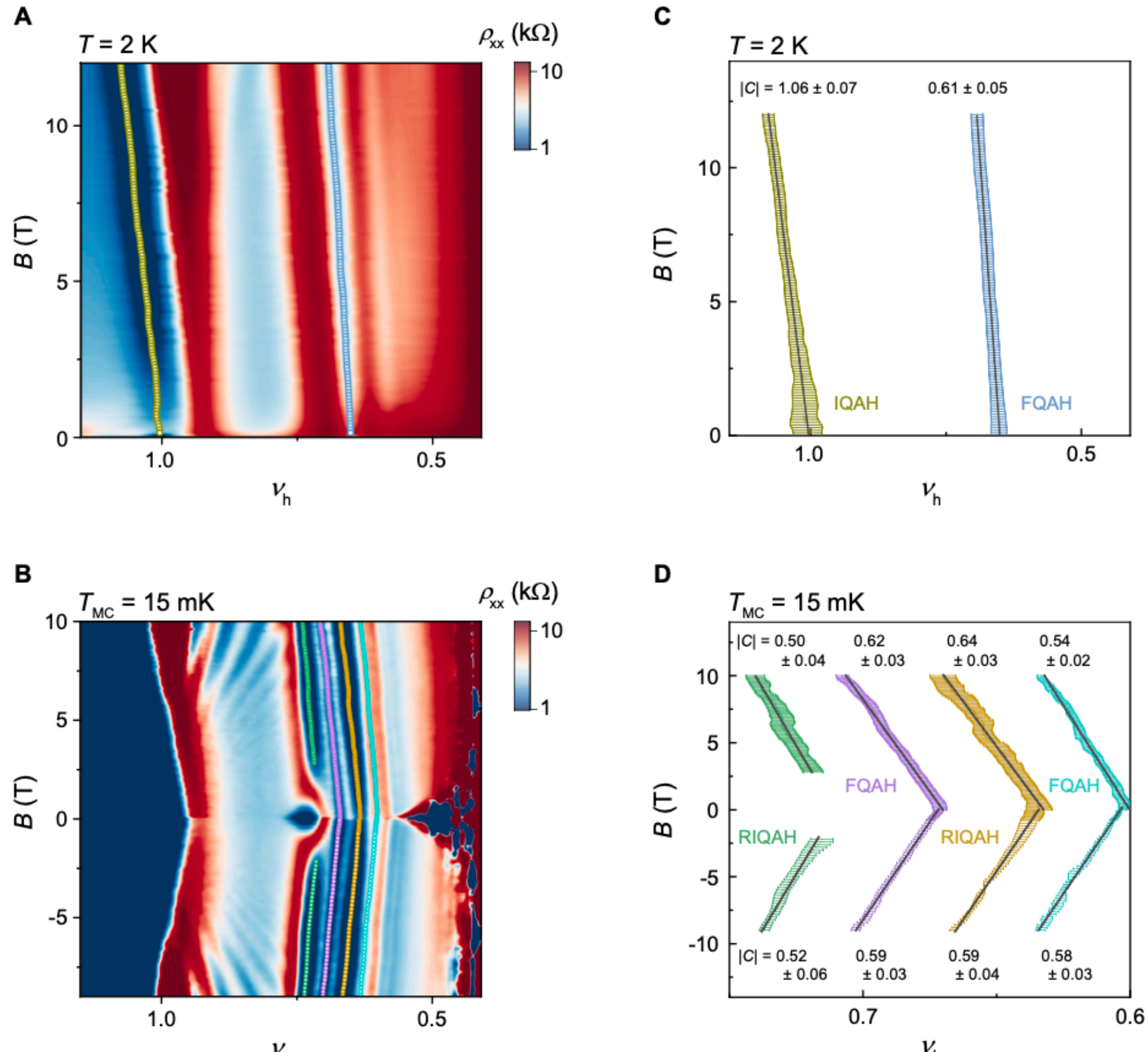

**Fig. S5. Chern number fitting of IQAH, FQAH and RIQAH states in Device 1.** (**A-B**) $\rho_{xx}$ as a function of $\nu_h$ and $B$ under $D = 0$ at $T = 2$ K (A) and $T_{MC} = 15$ mK (B), respectively. $\rho_{xx}$ at the IQAH, FQAH and RIQAH states exhibits local dips and disperses with $B$ and $\nu_h$. Different groups of colored circles represent the positions of ($\nu_h$, $B$) where $\rho_{xx}$ reaches its local minimum in the IQAH, FQAH and RIQAH states, respectively. (**C-D**) Black lines illustrate the linear fittings to the local $\rho_{xx}$ dips (with error bars) using the Streda formula of $n_M \frac{d\nu_h}{dB} = C\frac{e}{h}$, with the fitted Chern numbers labeled in the figures. Based on data at $T = 2$ K, the fitted Chen numbers for the $\nu_h = 1$ IQAH state and $\nu_h = 2/3$ FQAH state are $C = 1.06 \pm 0.07$ and $C = 0.61 \pm 0.05$, respectively. For $T_{MC} = 15$ mK, considering the fitting results at both positive and negative magnetic fields, we estimate the Chen numbers for the $\nu_h = 2/3$ and $3/5$ FQAH states to be $C = 0.61 \pm 0.05$ and $C = 0.56 \pm 0.05$. These values are in consistent with the measured quantized Hall conductance of the IQAH and FQAH states. However, for the RIQAH states, the observed dispersion does not align with the expected value of $C = 1$, as indicated by the measured $\rho_{xy}$ quantized at $h/e^2$.

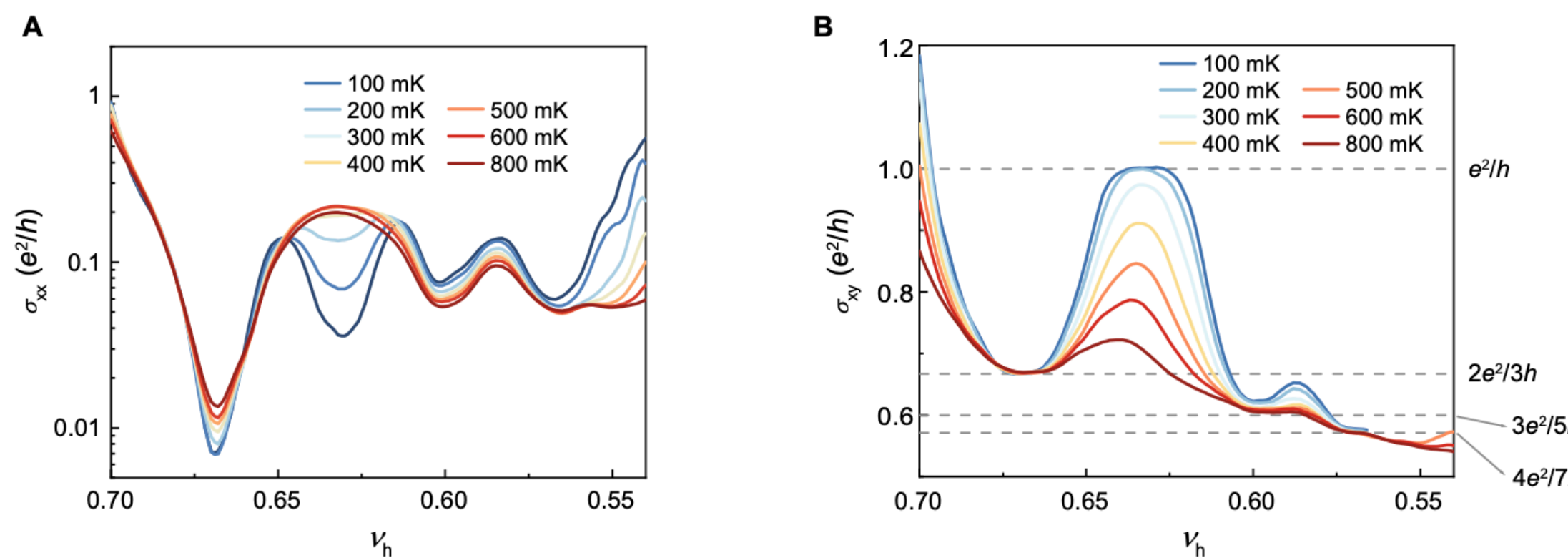

**Fig. S6. Temperature dependence of longitudinal conductivity and Hall conductivity in Device 1.** (**A-B**) Longitudinal conductivity $\sigma_{xx}$ (A) and Hall conductivity $\sigma_{xy}$ (B) versus $\nu_h$ at varying $T$ in Device 1. We derive $\sigma_{xx}$ and $\sigma_{xy}$ using the reciprocal resistance-to-conductance tensor conversion given by $\sigma_{xx} = \frac{\rho_{xx}}{\rho_{xx}^2+\rho_{xy}^2}$ and $\sigma_{xy} = \frac{\rho_{xy}}{\rho_{xx}^2+\rho_{xy}^2}$. Quantized $\sigma_{xy}$ plateaus for the $\nu_h$ = 2/3, 3/5, 4/7 FQAH states and the $\nu_h \approx$ 0.63 RIQAH state are observed. The quantization for $\nu_h$ = 3/5 and 4/7 FQAH states turns out to be better at relatively higher temperatures, which is unexpected. The precise origin and mechanism remain unclear and require further investigation.

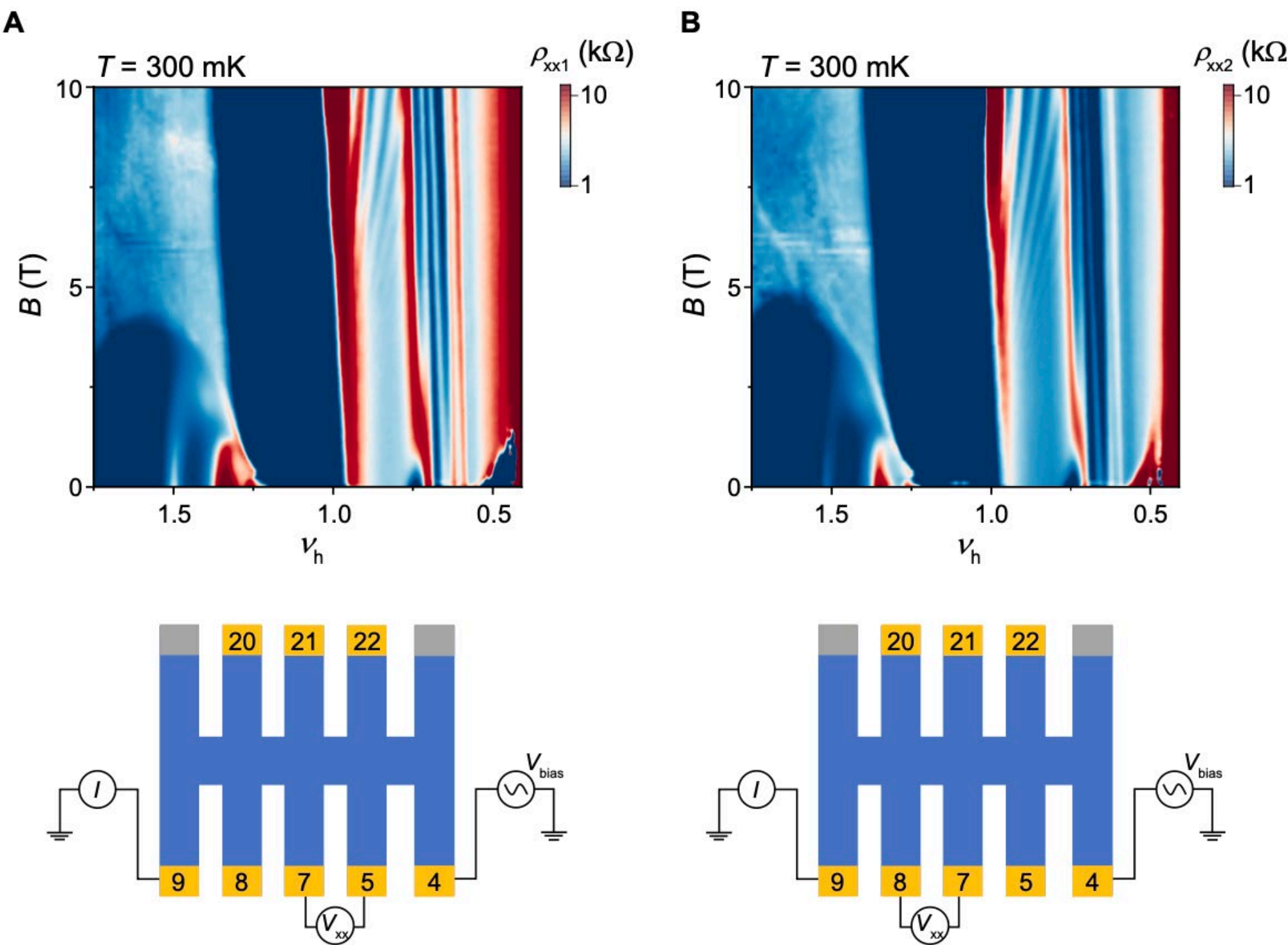

**Fig. S7. $\rho_{xx}$-$\nu_h$-$B$ maps measured with different contact pairs of Device 1.** (**A-B**) Symmetrized $\rho_{xx}$ as a function of $\nu_h$ and $B$ at $D$ = -5 mV/nm and $T$ = 300 mK, measured with different contact pairs. The lower panels show the corresponding measurement configurations. The measurement results obtained from the two configurations are in excellent agreement. The results of $\rho_{xx}$ shown in the main figures were mainly measured using the configuration shown in (A).

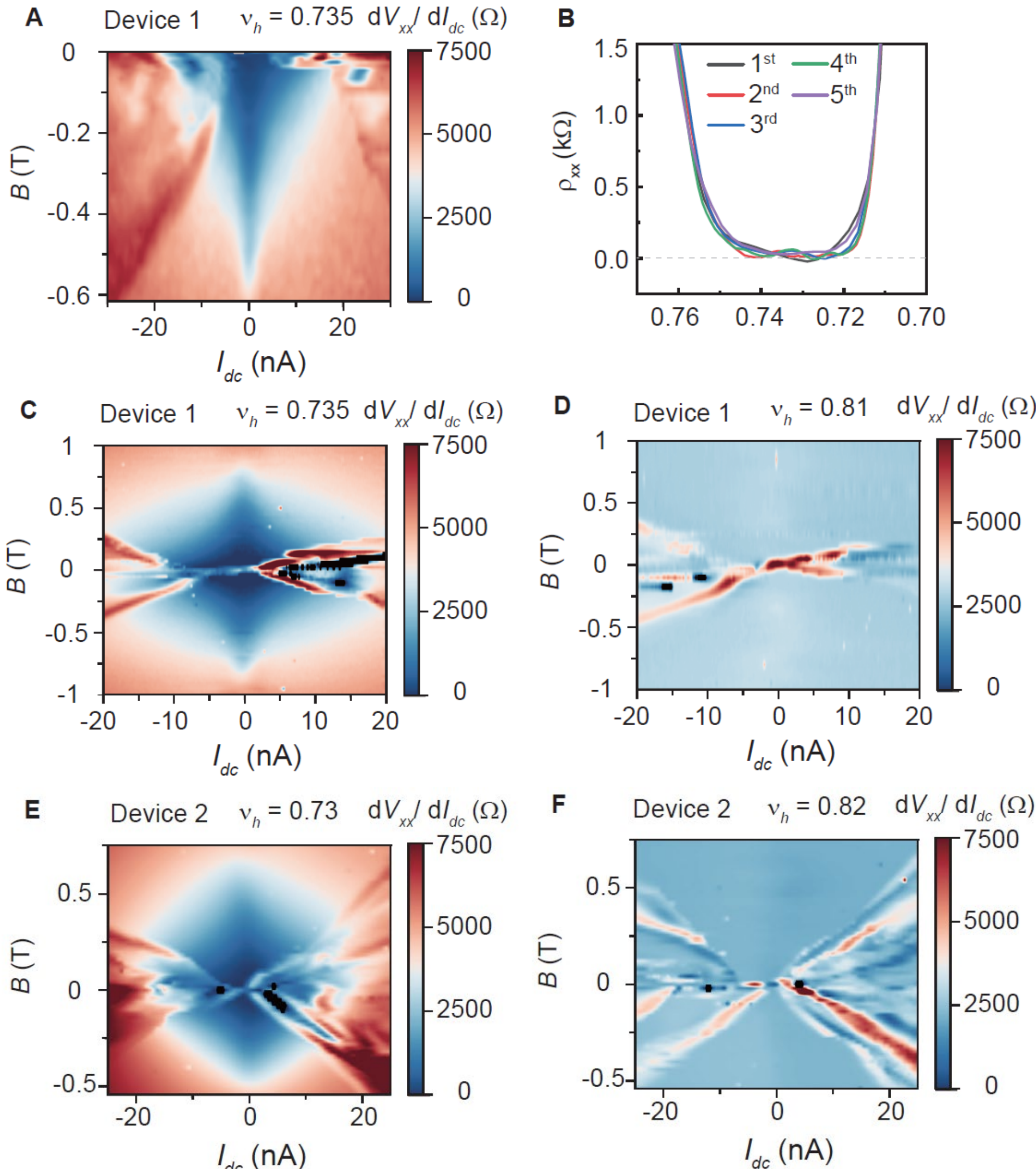

**Fig. S8. d*V*/d*I*-$I_{dc}$-*B* maps for the superconducting state.** (**A**), Differential resistance d*V*/d*I* as functions of dc current $I_{dc}$ and *B* measured in the top-loading dilution refrigerator (Oxford TLM) at *T* = 100 mK, $\nu_h$ = 0.735, and *D* = 0 in Device 1. (**B**), Multiple repeated measurements of $\rho_{xx}$ as a function of $\nu_h$ around the superconducting state in Device 1 at $T_{MC}$ = 15 mK and *B* = 0. A small residual resistivity of ~20-30 Ω with fluctuations near zero resistance is observed. This residual resistance is only about 0.5% of the corresponding superconducting normal-state resistivity, and also comparable with the measurement noise floor considering that the excitation current used in the measurements is only 1 nA. (**C**), The d*V*/d*I*-$I_{dc}$-*B* map of Device 1 measured in another cryogen free dilution refrigerator (Q-one, Q-400) at $T_{MC}$ = 10 mK, $\nu_h$ = 0.74, and *D* = 0. In both (A) and (C), we notice that, on top of the overall nonlinear *I*–*V* characteristics associated with superconductivity, additional peak and dip features appear in d*V*/d*I*. We tentatively attribute these features to magnetic domain switching, as they shift to higher currents with increasing *B*. Similar domain-switching behavior is also observed in the metallic state outside the superconducting dome (measured at $\nu_h$ = 0.81 and *D* = 0 at $T_{MC}$ = 10 mK, shown in **D**). Current-driven orbital magnetic domain switching in twisted $MoTe_2$ systems has also been recently reported (*78*). (**E-F**), Similar plots to those in (C) and (D), but measured in Device 2 at *T* = 300 mK. The main observations remain similar; however, more domain switch features are observed, presumably due to lower device quality compared with Device 1. The ac modulation current for all maps is 1 nA.

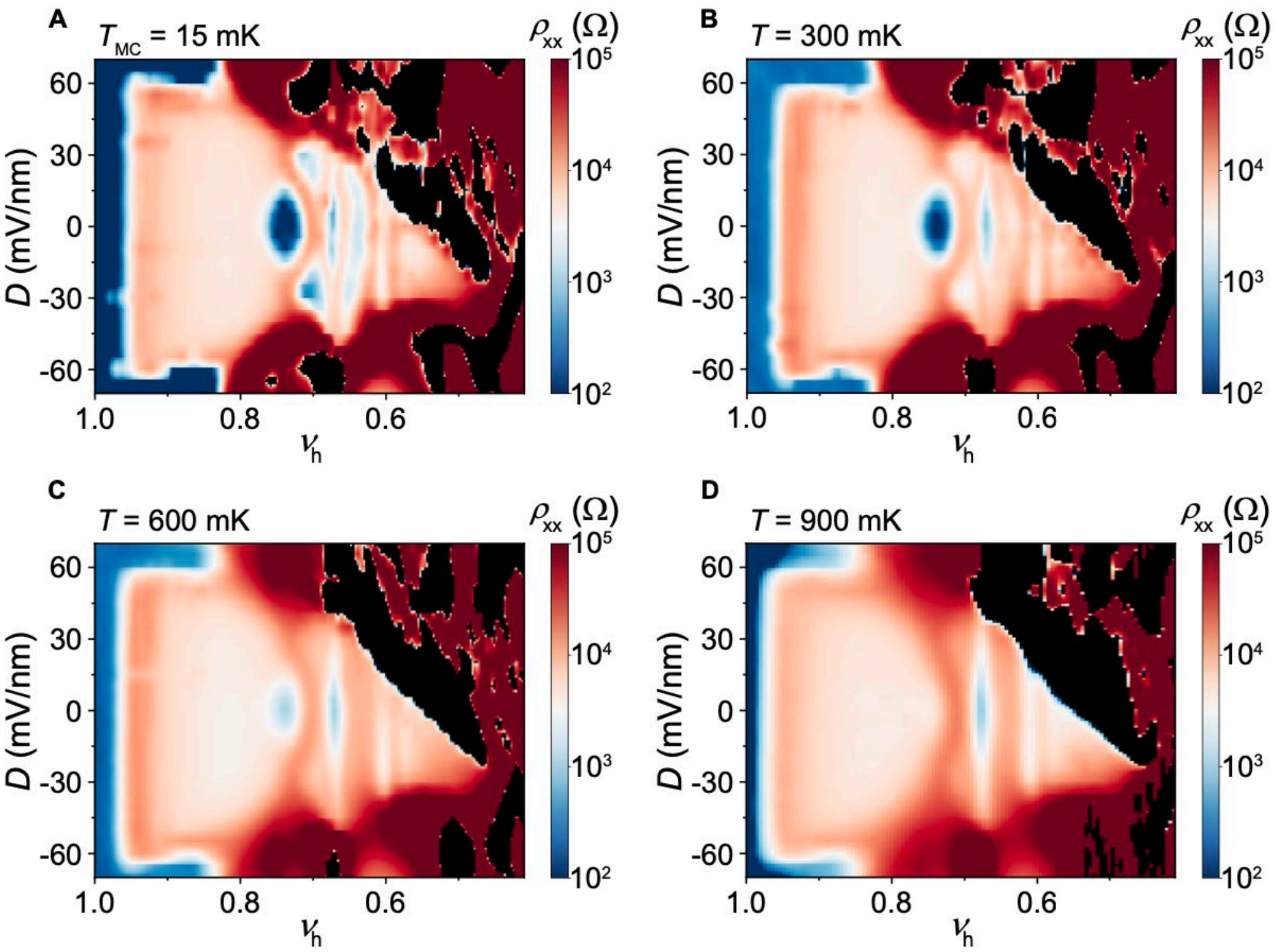

**Fig. S9. Temperature dependence of $\rho_{xx}$-$\nu_h$-$D$ maps at zero magnetic field.** (**A-D**) $\rho_{xx}$ as functions of $\nu_h$ and $D$, measured at $B = 0$ and $T_{MC}$ = 15 mK (A), 300 mK (B), 600 mK (C), and 900 mK (D) in Device 1. The black regions in (A-D) are experimentally inaccessible, either due to their highly insulating nature or contact issues. These maps clearly demonstrate that the superconducting state and RIQAH states possess smaller energy scale than the FQAH states in 3.83° $tMoTe_2$.

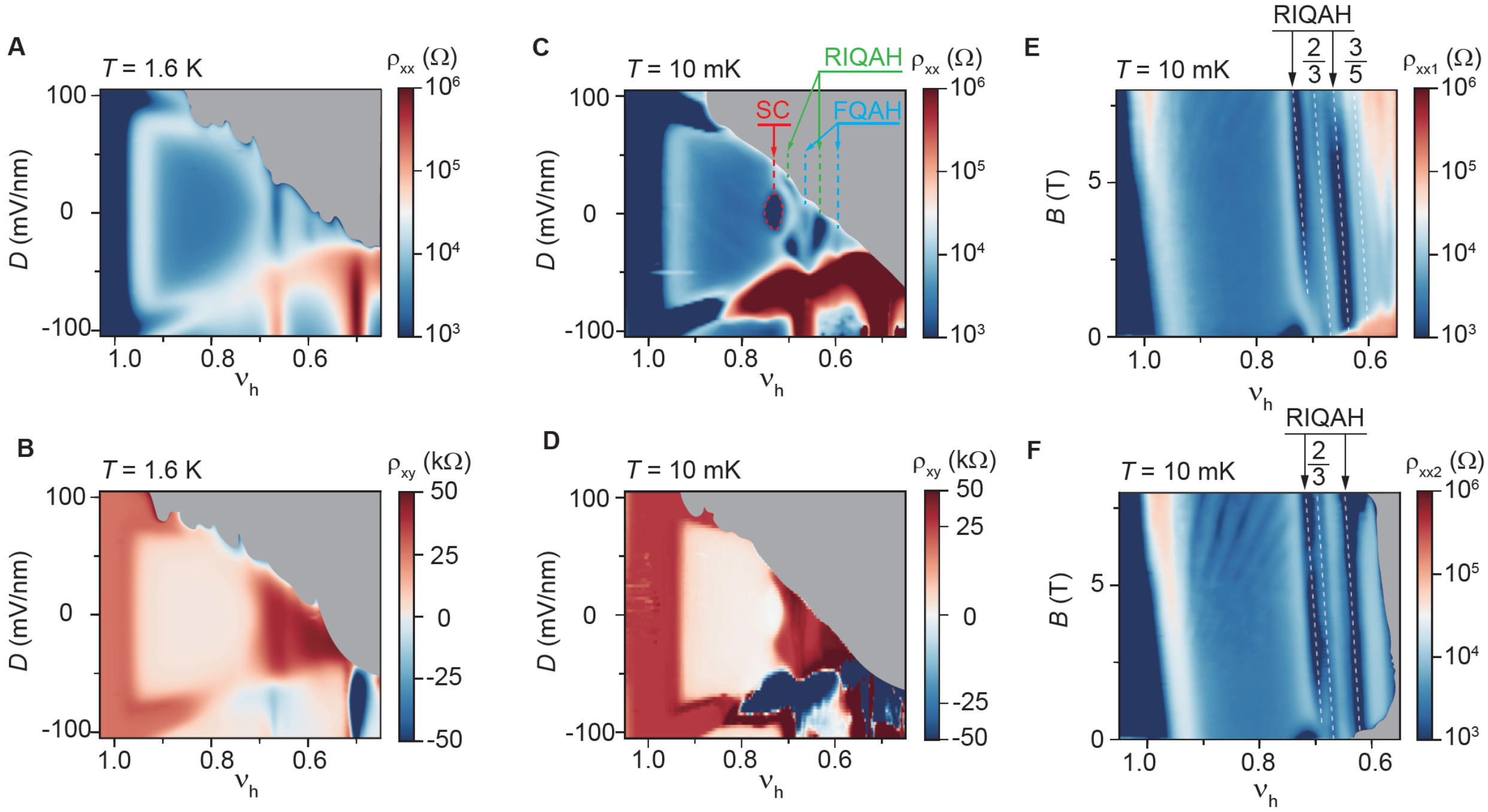

**Fig. S10. FQAH, RIQAH and superconducting states in Device 2.** (**A-B**) $\rho_{xx}$ (A) and $\rho_{xy}$ (B) as functions of $\nu_h$ and $D$, measured at $B$ = 0.1 T and $T$ = 1.6 K in Device 2 (3.89° $tMoTe_2$). (**C-D**) Symmetrized $\rho_{xx}$ (C) and anti-symmetrized $\rho_{xy}$ (D) as functions of $\nu_h$ and $D$, measured at $B$ = 0.1 T and $T_{MC}$ = 10 mK. FQAH, RIQAH, and superconducting features are highlighted in (C). (**E-F**) $\rho_{xx}$ as functions of $\nu_h$ and $B$, measured with different contact pairs in Device 2 at $D$ = 0 and $T_{MC}$ = 10 mK. The white dashed lines indicate the Středa dispersion with a Chern number $C$ equals to the corresponding fractional filling factors where the states appear. It can be seen that although the RIQAH states exhibit integer Hall quantization of $h/e^2$, the measured dispersion slopes are very close to their corresponding fractional filling factors rather than to $C$ = 1. These maps clearly demonstrate that the main observations, including the FQAH, RIQAH, and superconducting states in Device 2 are very similar to that observed in Device 1. The grey regions in maps are experimentally inaccessible.

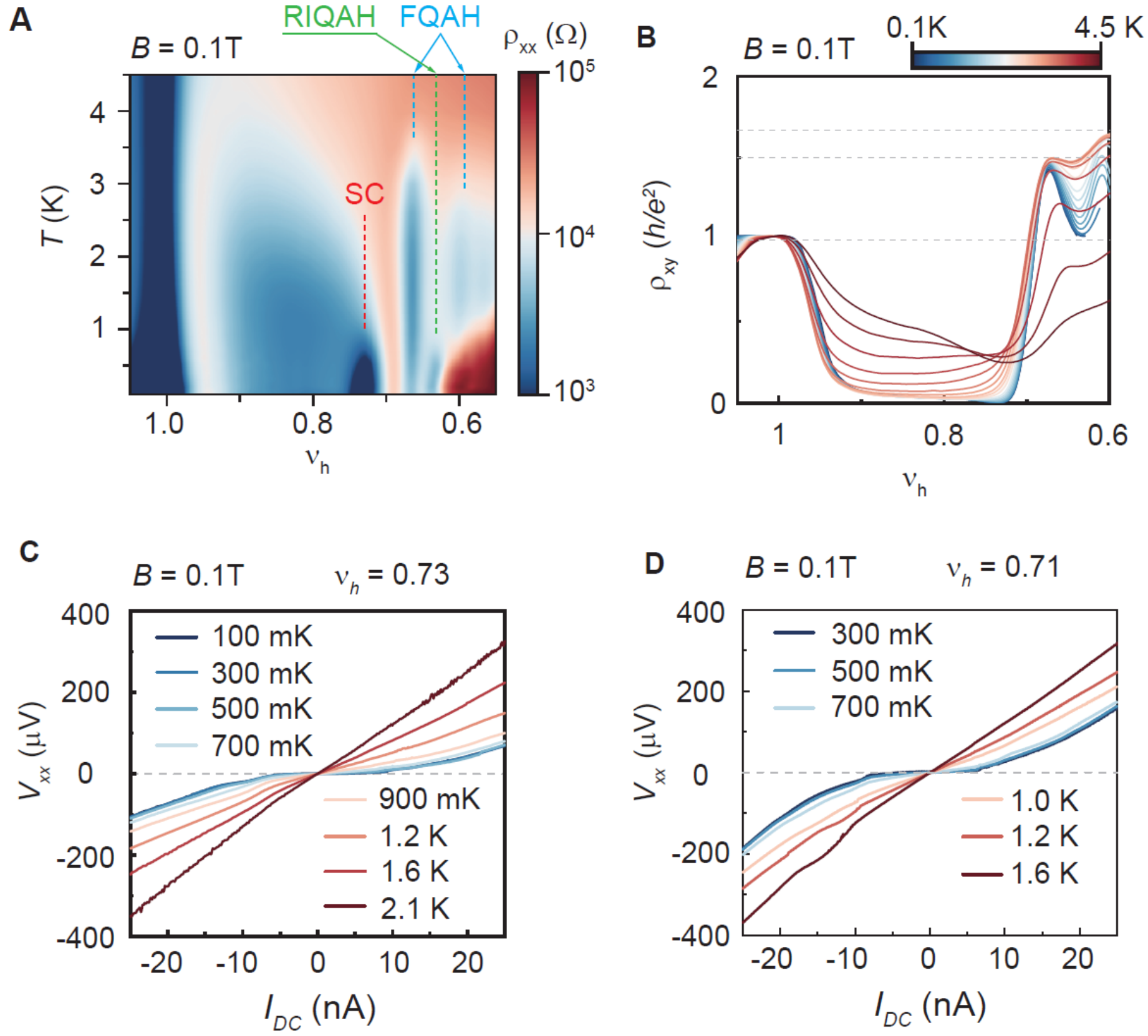

**Fig. S11. Temperature dependence and *I-V* curves of the superconducting states in Device 2.** (**A**) Symmetrized $\rho_{xx}$ as functions of $\nu_h$ and $T$, measured at $B = 0.1$ T and $D = 0$ in Device 2 (3.89° $tMoTe_2$). FQAH ($\nu_h = 2/3$ and $3/5$), RIQAH, and superconducting features are highlighted. (**B**) Anti-symmetrized $\rho_{xy}$ as functions of $\nu_h$, measured at $B = 0.1$ T, $D = 0$, and $T$ from 100 mK to 4.5 K in Device 2. (**C-D**) Temperature dependent *I-V* curves measured at $B = 0.1$ T within the superconducting state at $\nu_h = 0.73$ (C) and $\nu_h = 0.71$ (D). Clear nonlinear superconducting *I*–*V* characteristics are observed below ~ 1 K, with a critical current of approximately 5-10 nA. (C) and (D) were measured during two separate cooldowns.

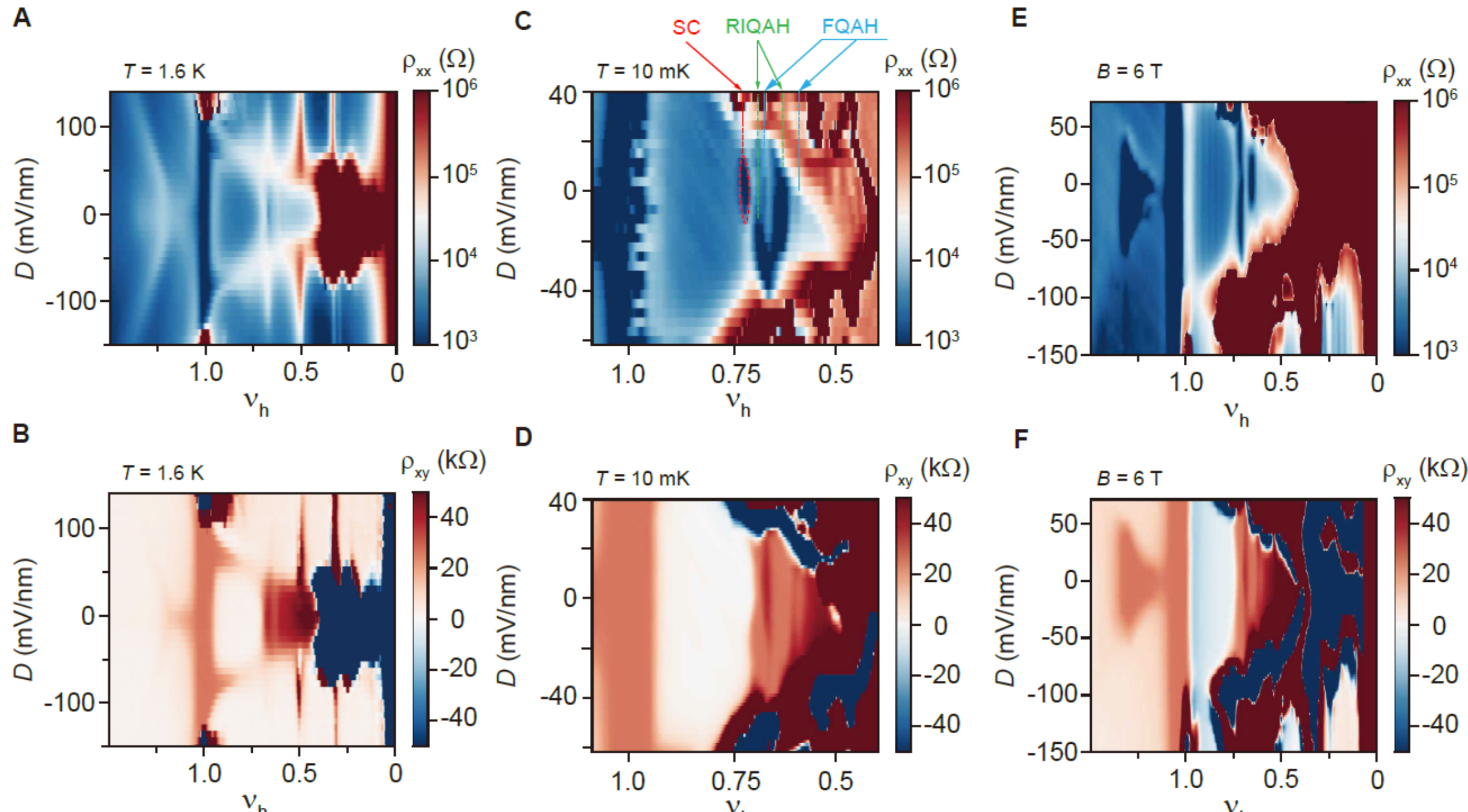

**Fig. S12. Transport $\nu_h$-$D$ phase diagrams of Device 3.** (**A-B**) $\rho_{xx}$ (A) and $\rho_{xy}$ (B) as functions of $\nu_h$ and $D$, measured at $B$ = 0.1 T and $T$ = 1.6 K in Device 3 (3.93° $tMoTe_2$). (**C-D**) Symmetrized $\rho_{xx}$ (C) and anti-symmetrized $\rho_{xy}$ (D) as functions of $\nu_h$ and $D$, measured at $B$ = 0.1 T and $T_{MC}$ = 10 mK. FQAH, RIQAH, and superconducting features are highlighted in (C). (**E-F**) $\rho_{xx}$ (E) and $\rho_{xy}$ (F) as functions of $\nu_h$ and $D$, measured at $B$ = 6 T and $T_{MC}$ = 10 mK. The phase diagrams observed in Device 3 are similar to that observed in Device 1 and 2, although the data quality is lower, likely due to stronger disorder effects in Device 3.

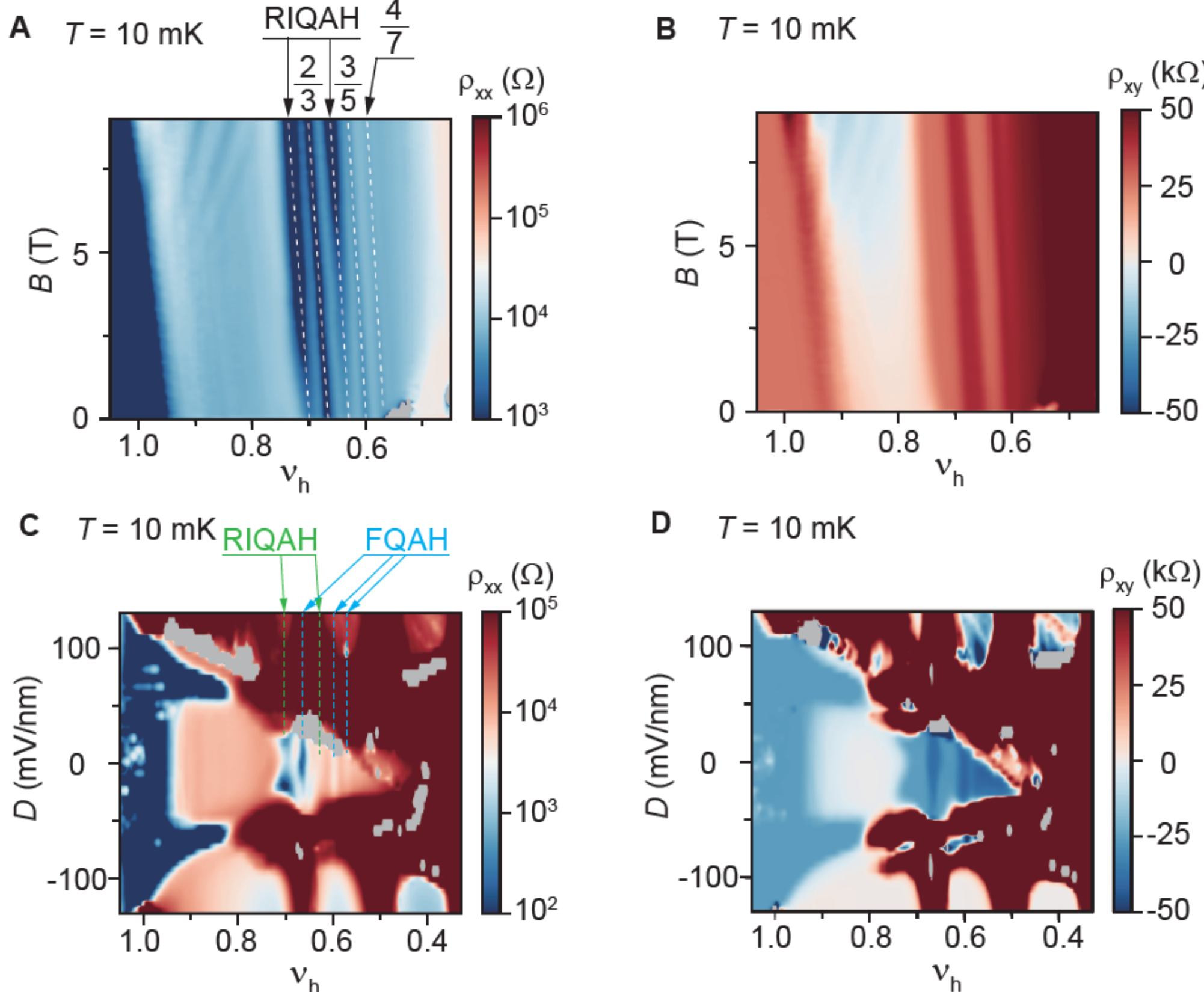

**Fig. S13. FQAH and RIQAH states in Device 4.** (**A-B**) $\rho_{xx}$ (A) and $\rho_{xy}$ (B) as functions of $\nu_h$ and $B$, measured in Device 4 (3.79° $tMoTe_2$) at $D = 0$ and $T_{MC} = 10$ mK. The white dashed lines indicate the Středa dispersion with a Chern number $C$ equals to the corresponding fractional filling factors where the states appear. Again, the RIQAH states are found to exhibit integer Hall quantization of $h/e^2$, but the measured dispersion slopes are closer to their corresponding fractional filling factors rather than to $C = 1$. (**C-D**) $\rho_{xx}$ (C) and $\rho_{xy}$ (D) as functions of $\nu_h$ and $D$, measured at $B = -0.1$ T and $T_{MC} = 10$ mK in Device 4. FQAH and RIQAH features are highlighted in (C). No superconducting states were observed in Device 4 despite the high device quality. The grey regions in maps are experimentally inaccessible.

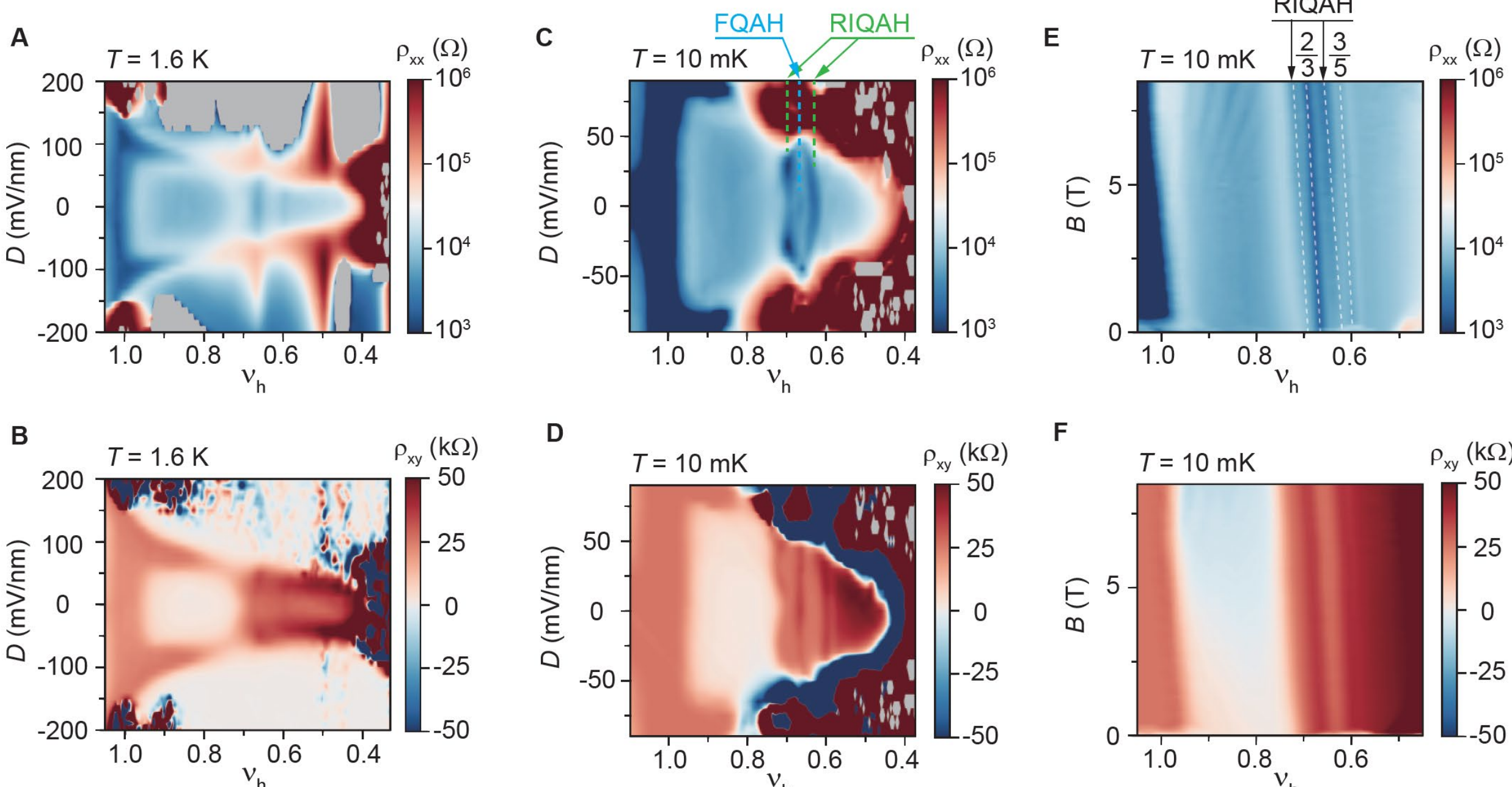

**Fig. S14. FQAH and RIQAH states in Device 5.** (**A-B**) Symmetrized $\rho_{xx}$ (A) and anti-symmetrized $\rho_{xy}$ (B) as functions of $\nu_h$ and $D$, measured at $B$ = 0.1 T and $T$ = 1.6 K in Device 5 (3.98° tMoTe$_2$). (**C-D**) Symmetrized $\rho_{xx}$ (C) and anti-symmetrized $\rho_{xy}$ (D) as functions of $\nu_h$ and $D$, measured at $B$ = 0.1 T and $T_{MC}$ = 10 mK. FQAH and RIQAH features are highlighted in (C). No superconducting states were observed in Device 5 despite the high device quality. (**E-F**) $\rho_{xx}$ (E) and $\rho_{xy}$ (F) as functions of $\nu_h$ and $B$ at $D$ = 0 and $T_{MC}$ = 10 mK. Similar to Fig. S10E, 10F and S13A, the white dashed lines indicate the Středa dispersion with a Chern number $C$ equals to the corresponding fractional filling factors where the states appear. Similar to the observation in Device 1-4, the RIQAH states does not follow the Středa dispersion with $C$ = 1. The grey regions in maps are experimentally inaccessible.

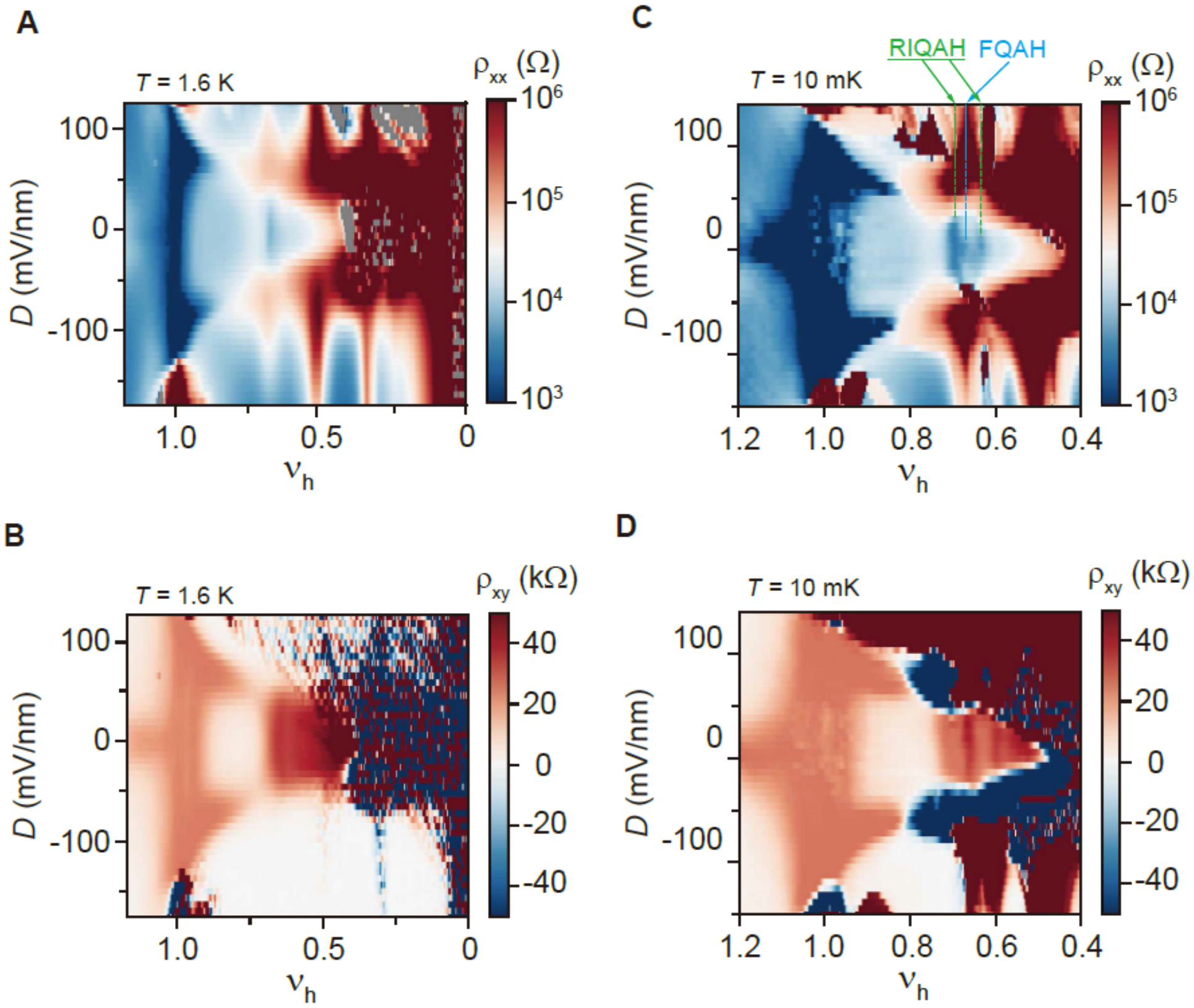

**Fig. S15. Transport $\nu_h$-$D$ phase diagrams of Device 6.** (**A-B**) Symmetrized $\rho_{xx}$ (A) and anti-symmetrized $\rho_{xy}$ (B) as functions of $\nu_h$ and $D$, measured at $B = 0.1$ T and $T = 1.6$ K in Device 6 (3.90° t$MoTe_2$). (**C-D**) Symmetrized $\rho_{xx}$ (C) and anti-symmetrized $\rho_{xy}$ (D) as functions of $\nu_h$ and $D$, measured at $B = 0.1$ T and $T_{MC}$ = 10 mK. FQAH and RIQAH features are highlighted in (C). No superconducting states were observed in this device, presumably due to relatively strong disorder effects compared with Device 1-3. The grey regions in maps are experimentally inaccessible.

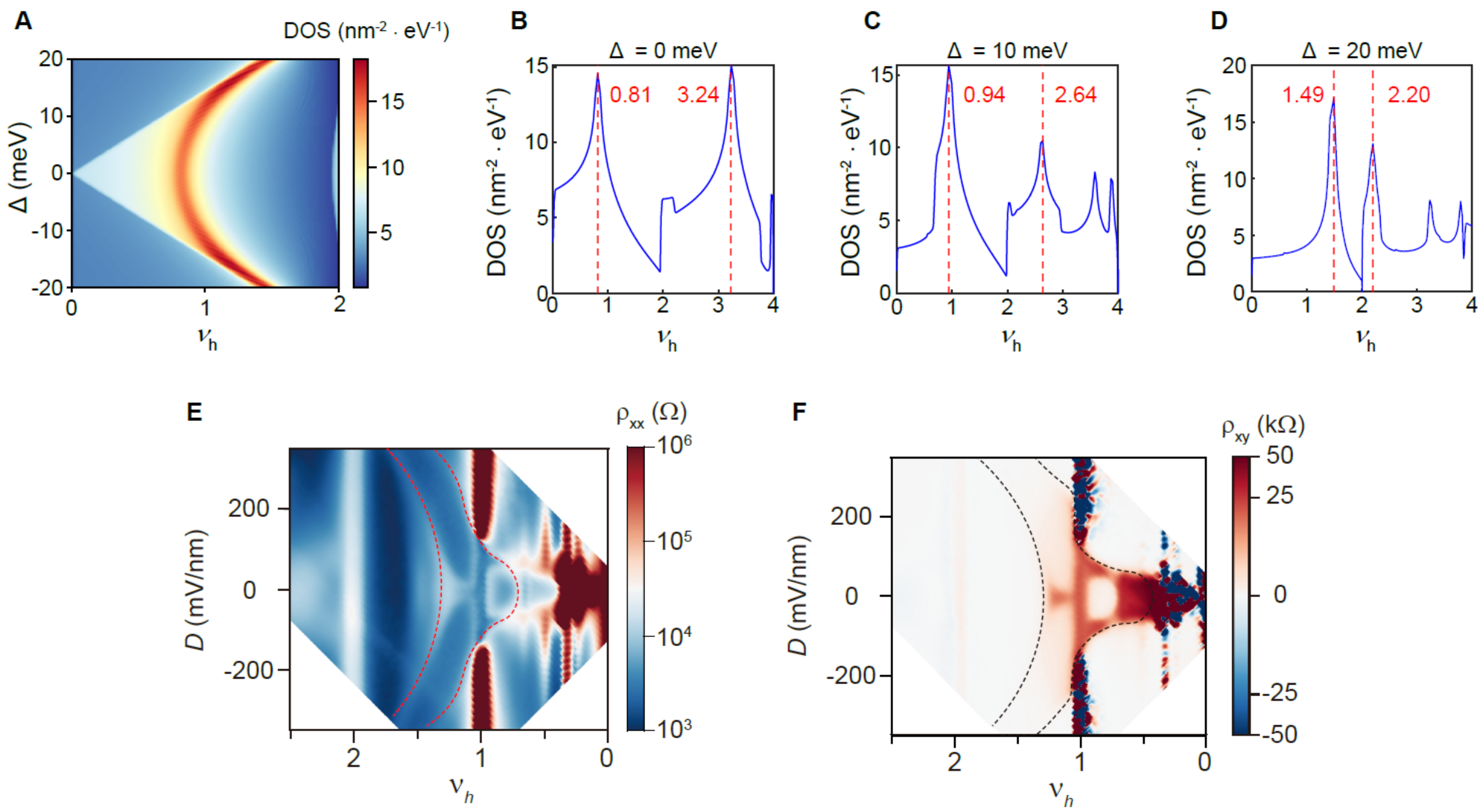

**Fig. S16. Van Hove singularities in tMoTe₂.** (**A**) Calculated electronic density of states (DOS) as a function of sublattice potential difference Δ (the energy difference induced by the displacement field $D$ between layers) and $\nu_h$. The van Hove singularity (VHS), located at $\nu_h \approx 0.80$ when $\Delta = 0$ (assuming valley degeneracy), shifts toward higher values of $\nu_h$ as the Δ increases. (**B-D**), DOS at $\Delta = 0$ meV (B), 10 meV (C), and 20 meV (D). The VHS is highlighted by dashed lines. The conversion between Δ and $D$ can be estimated by $\Delta = eDt/\varepsilon_{\mathrm{MoTe_2}}$. Here $t \approx 0.7$ nm is the interlayer separation between the $MoTe_2$ monolayers. The out-of-plane dielectric constant of bilayer $MoTe_2$ (*79*) is about $\varepsilon_{\mathrm{MoTe_2}} \approx 10$. Based on the above estimations, $\Delta = 10$ meV corresponds to $D \approx 140$ mV/nm. (**E-F**), Symmetrized $\rho_{xx}$ (E) and anti-symmetrized $\rho_{xy}$ (F) as functions of $\nu_h$ and $D$, measured at $B = 0.3$ T and $T = 2$ K for device 1. Two branches of VHS are marked by red dashed lines in (A). Dashed lines in (B) represent the ferromagnetic region boundary. The observed VHS features deviate substantially from the single-particle calculation results shown in (A-D). In particular, two parabola-shaped VHS branches are resolved near $\nu_h \approx 0.7$ and $\nu_h \approx 1.3$ at $D = 0$, whose envelope closely follows the boundary of the zero-field anomalous Hall (ferromagnetic) region. This splitting can be naturally explained by interaction-induced valley polarization, and the coincidence of the two VHS branches with the ferromagnetic boundary suggests that the spontaneous valley polarization is closely connected to the diverging density of states associated with the VHS. It is also worth noting that interaction effects can strongly renormalize the band structure, making it difficult to quantitatively determine the exact shape and filling position of the VHS in the valley-polarized regime from theoretical calculations. So, based on the present transport measurements alone, we cannot unambiguously determine whether the system is fully valley polarized at the filling factors where the superconducting states emerge. Further experimental studies, such as using nano-SQUID or magneto-optical probes, will be helpful in clarifying this point.